\documentclass[journal]{IEEEtran}
\usepackage{lipsum}
\usepackage{mathtools}
\usepackage{cuted}

\usepackage{amsmath}
\usepackage{setspace}
\usepackage{flushend, cuted}
\usepackage{lingmacros}
\usepackage{tree-dvips}
\usepackage{varioref}
\usepackage{colortbl}
\usepackage{graphicx}
\usepackage{epstopdf}
\usepackage{color,soul}
\usepackage{amsfonts}
\usepackage{subfig}
\usepackage{multicol}
\usepackage{multirow}
\usepackage{hhline}
\usepackage{amssymb}
\usepackage{newlfont}
\usepackage{mathrsfs}
\usepackage{mathrsfs}
\usepackage{cite}
\usepackage{multirow}
\usepackage{nomencl}   
\usepackage{fancyhdr}
\usepackage{caption}
\usepackage{slashbox}
\usepackage{float}
\usepackage{placeins}
\usepackage{tabularx}
\usepackage[table]{xcolor}
\usepackage{helvet}
\usepackage{xcolor}

\ifCLASSINFOpdf

\else
 
\fi

\begin{document}

\title{Virtual Source Synthetic Aperture \\ for Accurate Lateral Displacement Estimation\\ in Ultrasound Elastography }
\author{Morteza~Mirzaei,
       Amir~Asif
        and~Hassan~Rivaz%
\thanks{Morteza Mirzaei, Amir Asif and Hassan Rivaz are with the Department
of Electrical and Computer Engineering, Concordia University, Montreal,
QC, H3G 1M8, Canada.
 Email: m\_irzae@encs.concordia.ca~,  amir.asif@concordia.ca ~and~~hrivaz@ece.concordia.ca}}

\maketitle

\begin{abstract}
Ultrasound elastography is an emerging non-invasive imaging technique wherein pathological alterations can be visualized by revealing the mechanical properties of the tissue.  Estimating tissue displacement in all directions is required to accurately estimate the mechanical properties.  Despite capabilities of elastography techniques in estimating displacement in both axial and lateral directions, estimation of axial displacement is more accurate than lateral direction due to higher sampling frequency, higher resolution and having a carrier signal propagating in the axial direction. Among different ultrasound imaging techniques, Synthetic Aperture (SA) has better lateral resolution than others, but it is not commonly used for ultrasound elastography due to its limitation in imaging depth of field. 
Virtual source synthetic aperture (VSSA) imaging is a technique to implement synthetic aperture beamforming on the focused transmitted data to overcome limitation of SA in depth of field while maintaining the same lateral resolution as SA. Besides lateral resolution, VSSA has the capability of increasing sampling frequency in the lateral direction without interpolation. 
In this paper, we utilize VSSA to perform beamforming to enable higher resolution and sampling frequency in the lateral direction. The beamformed data is then processed using our recently published elastography technique, OVERWIND~\cite{overwind}. Simulation and experimental results show substantial improvement in estimation of lateral displacements.                                 
                
\end{abstract}


%
\IEEEpeerreviewmaketitle
                  
\section{Introduction}
Ultrasound elastography (USE) is a technique for detecting
alterations in mechanical properties of tissue using ultrasound imaging, which is a widely available modality and offers the additional advantage of being non-invasive and low cost.
As such, USE may help in early diagnosis and improves the prognosis of treatments.  In recent years,  USE has been utilized in several
clinical applications  including ablation guidance
and monitoring \cite{sigrist2017ultrasound}, differentiating
benign thyroid nodules from malignant ones \cite{hong2009real,trimboli2012ultrasound,samir2015shear} and breast lesion characterization \cite{hall2001vivo,doyley2001freehand,uniyal2015ultrasound}. Surgical treatment of liver cancer \cite{rivaz2014ultrasound,yang2014monitoring,liverlas}, assessment
of fibrosis in chronic liver diseases \cite{tang2015ultrasound, tsochatzis2011elastography}, 
detecting prostate cancer \cite{lorenz1999new, correas2013update}, differentiating abnormal
lymph nodes in benign conditions \cite{saftoiu2006endoscopic} and brain tumor
surgery \cite{selbekk2005strain, selbekk2010tissue} are other relevant clinical applications of USE.

Pathological alterations are correlated with the mechanical properties of tissue, and for each material, 81 constants are required to describe the fourth-order stiffness tensor \cite{theoryofelasticity,ophir1999elastography}. Many of these parameters depend on each other in a valid stiffness tensor. Moreover, for most clinical applications, the tissue can be assumed linear elastic and isotropic, which reduces the number of parameters to 2 independent ones  \cite{ophir1999elastography}. These two constants are known as the Lam\'e  constants, or their equivalents,  Young's modulus and Poisson's ratio. 

Different methods are proposed for estimating the Young's modulus and Poisson's ratio, which  can be broadly grouped into dynamic and quasi-static elastography. Dynamic methods, such as shear wave imaging (SWI)~\cite{bercoff2004supersonic} and acoustic radiation force imaging (ARFI) \cite{nightingale2003shear,dumont2015robust}, use Acoustic Radiation Force (ARF) to stimulate displacement in tissue. Quasi-static elastography proposed in \cite{ophir1991elastography} use external excitation by slowly pressing the probe against the tissue  utilizing a robotic arm \cite{schneider2012remote,adebar2011robotic} or a hand-held probe (i.e. free-hand palpation)  \cite{xia2014dynamic,hall2003vivo}. Even though the induced compression is uni-axial, tissue deforms in all directions due to its incompressible property where the volume of the tissue does not change when compressed. The first step of quasi static elastography is time delay estimation (TDE), wherein tissue deformation should be estimated and differentiated to provide the strains.  In the next step the inverse problem should be solved to estimate the Young's modulus based on
strains \cite{skovoroda1995tissue,pan2014regularization}.

\begin{figure*}
	\centering
	\subfloat[SA]{{\includegraphics[height=5cm]{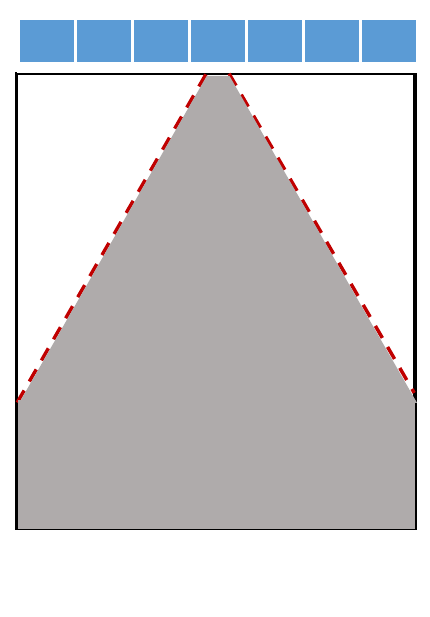}}}
	\quad
	\subfloat[Line by line]{{\includegraphics[height=5cm]{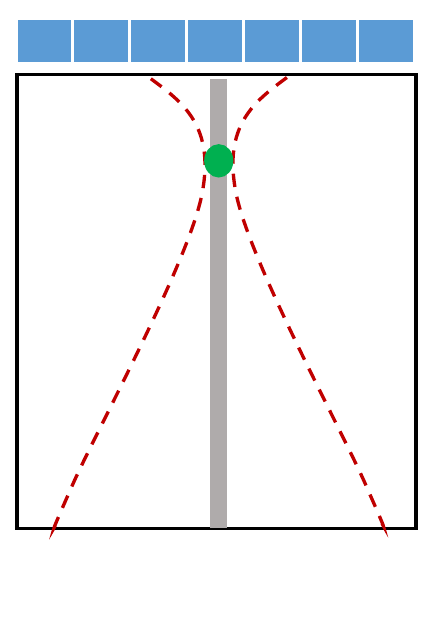}}}
	\quad
	\subfloat[VSSA]{{\includegraphics[height=5cm]{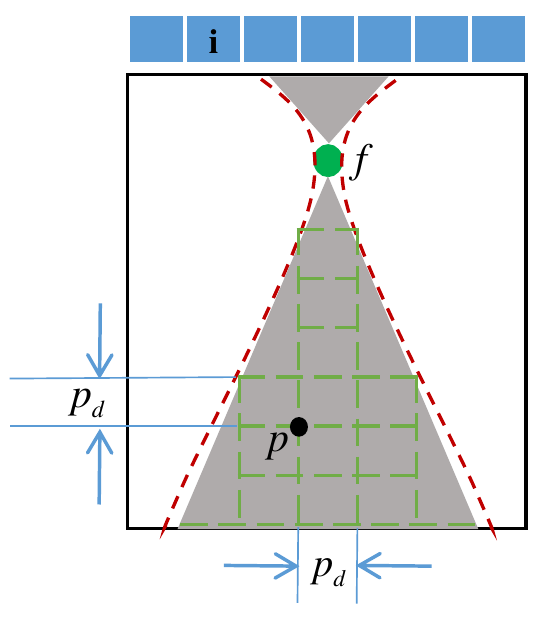}}}	
	\caption{Schematics of different imaging modes.  The red dashed lines show the beam pattern, while the gray areas show the regions that received data should be focused. (a) shows the SA imaging in which a wave propagates in tissue. (b) is line by line imaging that transmitted beam gets narrow at focal point, which is shown by a green circle. (c) is VSSA in which the transmission is similar to line by line imaging and the focus point is assumed as virtual source of transmission.}
	\label{imaging}
\end{figure*} 
Different methods are proposed to perform TDE that can be broadly categorized as  window-based, regularized optimization-based, and deep-learning approaches \cite{tehrani2020displacement,tehrani2020semi}.
For estimating the displacement of a sample, the window based approaches consider a window around each sample and estimate the displacement of each window by determining a window closest in pixel values in the next frame. Several similarity metrics are proposed to compare the windows of pre-compressed and post-compressed tissue such as normalized cross correlation (NCC) of windows~\cite{varghese2000direct,mirzaei20203d,zahiri2006motion}, phase-correlation wherein zero crossing of phase determines displacement \cite{yuan2015analytical} and sum of absolute difference of windows~\cite{chaturvedi1998testing}.  
Another class of TDE methods is  regularized optimizing based technique that imposes regularization between neighboring samples \cite{rivaz2011real,mirzaei2020accurate}. We have recently proposed a method called OVERWIND, which is a combination of window-based approaches and regularized optimization method to take advantages of both methods.

Despite the capability of OVERWIND in estimating both axial and lateral displacements, the latter is of lower quality compared to the former for three main reasons: low sampling rate, lack of carrier signal and low resolution in the lateral direction \cite{luo2009effects,he2017performance}.  One of the most utilized techniques for increasing the data size in lateral direction is interpolation \cite{konofagou1998new,liu2017systematic}.  It is shown by experimental results that spline interpolation has the best performance among different techniques of interpolation \cite{luo2009effects}. 
In these methods, the bandwidth should be large enough to have sufficient overlap between adjacent lines \cite{konofagou1998new}. It is shown that the minimum density of A-lines for original data should be at least 2 A-line per beam width to have an acceptable interpolation \cite{konofagou1998new}. Accordingly, not only interpolation does not change the resolution, it cannot be implemented for high resolution data to increase the number of samples.
Besides low resolution, another disadvantage of interpolation data especially for large factor interpolations is decreasing the robustness of the TDE, since interpolation can be a source of error \cite{ebbini2006phase}.

Synthetic Aperture (SA) imaging is used for lateral strain estimation in \cite{korukonda2011estimating}. 
	SA has narrow and fixed beam width in all field of view in contrast to line by line imaging, which has narrow beam width in the focal zone only. Moreover, SA has capability of increasing sampling rate in lateral direction without interpolation, which also improves the resolution.
	It is shown that the accuracy of TDE increases by decreasing  the beam width \cite{korukonda2011estimating}. Accordingly, SA is better than line by line imaging for lateral elastography, with the disadvantage of lower transmission power and penetration depth, which could hinder clinical use of SA  \cite{korukonda2013noninvasive,nayak2017principal}. 
  
In this paper, we propose to use Virtual Source Synthetic Aperture (VSSA) imaging that implements SA-based beamforming on focused transmitted signals. On the one hand, this enables us to benefit from advantages of SA such as high resolution and the capability to increase the sampling frequency to increase the resolution and number of A-lines. On the other hand, we can take advantageous of line by line imaging in high penetration depth. Then the beamformed data is fed to our recently published TDE method, OVERWIND \cite{overwind} that has shown to outperform window-based and other regularized optimization-based techniques. We call the results of OVERWIND on VSSA with high sampling frequency in lateral direction as High Frequency OVERWIND (HF OVERWIND) and compare the results with OVERWIND on spline based interpolated data (Inter. OVERWIND) and also with OVERWIND on  VSSA with low sampling frequency in lateral direction in which the number of A-lines is equal to the number of piezo-electrics (LF OVERWIND). 
\section{METHODS}
Most elastography techniques like OVERWIND requires two sets of data
collected as the tissue undergoes some deformation.  Let $I_{1}$ and $I_{2}$ of size $(m,n)$ be the  beamformed RF data where $m$ and $n$ are depth and width of the imaged tissue. The goal of TDE is estimating the displacement field between these two data sets. 
In this section, we first briefly review our 
recently developed ultrasound elastography method, OVERWIND~\cite{overwind}, and then present the beamforming technique to increase the number of lines and resolution in the lateral direction to help OVERWIND in accurately estimating displacements.

\subsection{OVERWIND: tOtal Variation Regularization and WINDow-based time delay estimation}
The displacement estimation in OVERWIND comprises two steps for increasing the capabilities of the technique in estimating large deformations. In the first step, an integer estimation of the displacement is calculated using Dynamic Programming (DP), which is a recursive optimization based method for image registration. In this method, we consider a range of displacements for each sample and optimize the cost function that incorporates similarity of RF samples and displacement continuity to estimate integer displacement of RF samples \cite{rivaz2008ultrasound}. In the second step, the sub-sample displacements are calculated by minimizing the following cost function:  

\begin{equation*}
	\begin{array}{l}
	C(\varDelta a_{1,1},\ldots,\varDelta l_{m,n})= \varSigma_{j=1}^{n}\varSigma_{i=1}^{m}
		\bigg[\frac{1}{L}\varSigma_{k,r}\Big(I_1(i+k,j+r)\\ -I_2(.)-\varDelta a_{i,j} I'_{2a}(.) 
		-\varDelta l_{i,j} I'_{2l}(.)\Big)^2 \\
		+\alpha_1 \delta_1(a_{i,j}+\varDelta a_{i,j}-a_{i-1,j}-\varDelta a_{i-1,j}-\varepsilon_a) \\
		+ \alpha_2 \delta_2 (a_{i,j}+\varDelta a_{i,j}-a_{i,j-1}-\varDelta a_{i,j-1})\\
		+  \beta_1  \delta_3 (l_{i,j}+\varDelta l_{i,j}-l_{i-1,j}-\varDelta l_{i-1,j}) \\
		+ \beta_2 \delta_4 (l_{i,j}+\varDelta l_{i,j}-l_{i,j-1}-\varDelta l_{i,j-1}-\varepsilon_l)\bigg],
	\end{array}
	\label{costoverwind}
\end{equation*}  
where $i$ and $j$ are indices of RF samples in the region of interest and the symbols $i+k$ and $j+r$ represent indices of RF samples  inside the window that is considered around each sample. $\{a_{i,j}$, $\varDelta a_{i,j}\}$ and $\{l_{i,j}$,$\varDelta l_{i,j}\}$ represent the integer and sub-sample displacement of $(i,j)$ in axial and lateral directions, respectively.  $I_2(.)$ represent $I_2(i+k+a_{i,j},j+r+l_{i,j})$ and $I'_{2a}(.)$ and $I'_{2l}(.)$ are derivatives of $I_2$ in axial and lateral directions, respectively. $\delta_x(s)=2\lambda_x \sqrt{\lambda_x^2+s^2}$ is an approximate of norm L1 for regularization which allows sharp transitions where $\lambda_x$ is a scaling parameter.
Finally, $\alpha_1, \alpha_2, \beta_1$ and $\beta_2$ are regularization parameters to be tuned. These four parameters can be related to each other as explained in Discussion Section. 

OVERWIND considers both displacements in axial and lateral directions, but the estimation in the former direction is more accurate since, among other reasons, ultrasound data usually has less samples in lateral direction compared to axial direction. One of the most common techniques to cope with this issue is interpolating data in lateral direction. Not only it does not change the resolution but also its performance can deteriorate for the high resolution data sets since the A-lines do not have high overlap with each other in the high resolution data \cite{luo2009fundamental,luo2009key}.  Another disadvantage of interpolation is low robustness for complex interpolation techniques.
  
In this paper, we propose to use VSSA imaging mode for ultrasound elastography, which has the ability of increasing sampling frequency in the lateral direction as much as axial direction and also has high resolution in lateral direction while allowing high penetration. In the next subsection we describe SA, line by line imaging, and then show how VSSA can benefit  to relax two main limitations for lateral displacement estimation in ultrasound elastography.     

\subsection{SA: Synthetic Aperture}
In SA, a single element transmits a wave through the tissue as shown in Fig. \ref{imaging}(a)  and all elements record reflections. Each element generates an image of tissue (the gray area of Fig. \ref{imaging}(a)) by focusing the received beam at any point according to the expression
\begin{equation}
t_p({ij})=\dfrac{\sqrt{(x_p-x_i)^2+(z_p)^2}+\sqrt{(x_p-x_j)^2+(z_p)^2}}{c}
\end{equation}
where $c$ is the speed of sound in soft tissue and $i$ and $j$ are the transmitter and receiver elements symbols. $x_i$, $x_j$ and $x_p$ are the horizontal positions of the transmitter $i$, receiver $j$ and point $p$ where the beam is focused and $z_p$ is depth of point $p$ by assuming the probe is at zero depth. During each transmission, all receivers in the aperture focus the received beam at all points of the aperture  and summation of these data for all receivers generate a low-resolution image. The next element of the array transmits and the previously described operation is repeated to generate another low-resolution image. By repeating the experiment for all piezo-electrics as transmitter and adding up all low-resolution images, the final image is generated as per the following expression
 \begin{equation}
y_p=\sum\limits_{i=1}^e \sum\limits_{j=1}^e t_p({ij}),
\end{equation}
where $e$ is the total number of elements in the transducer array.
The main disadvantage of this technique for ultrasound elastography is the limitation in imaging deep areas since the emitted signal from one piezo-electric does not have enough power to penetrate deep areas.
\begin{figure*}
	\centering
	\subfloat[Ground truth]{{\includegraphics[width=4cm]{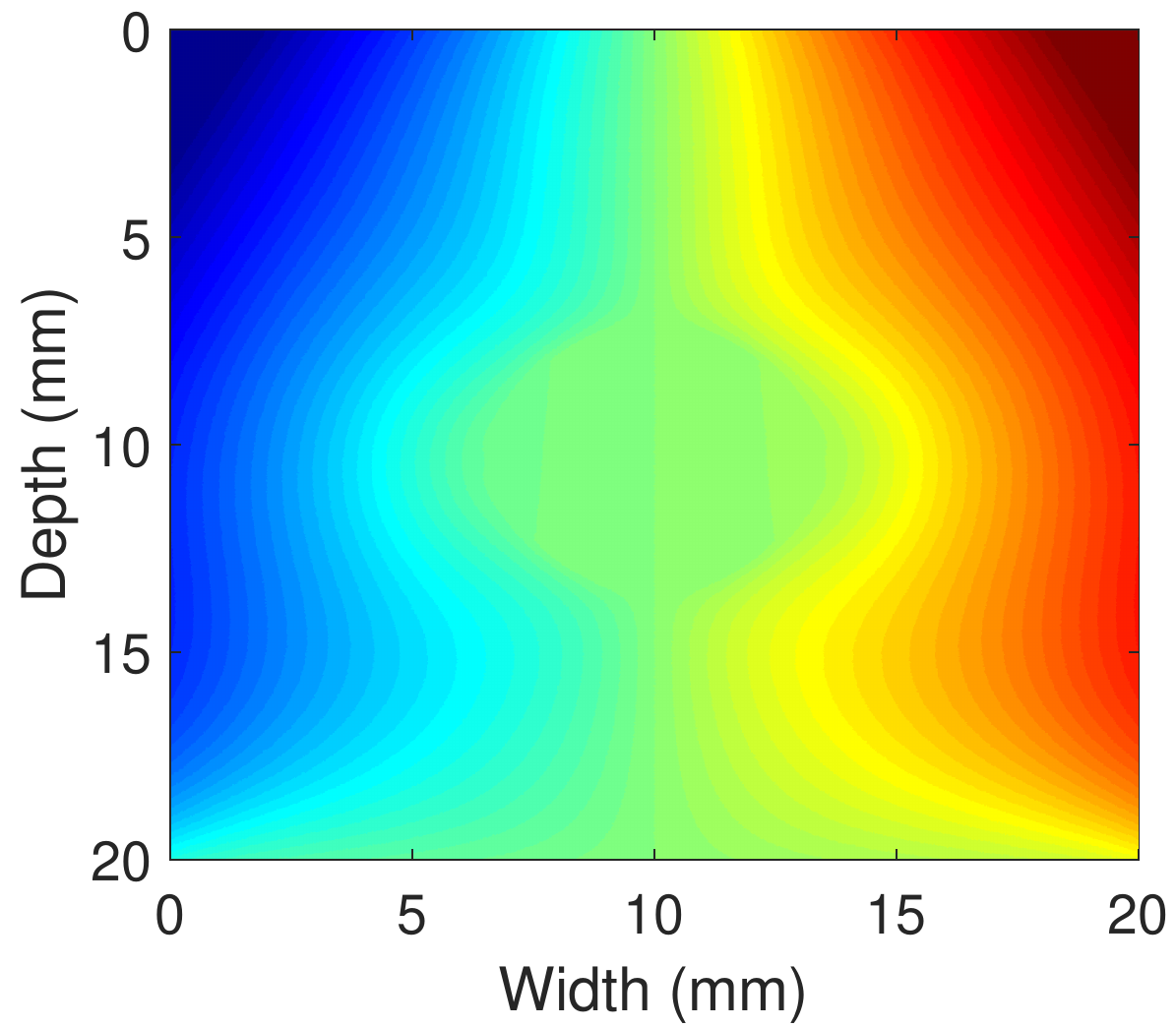}}}
	\subfloat[LF OVERWIND]{{\includegraphics[width=4cm]{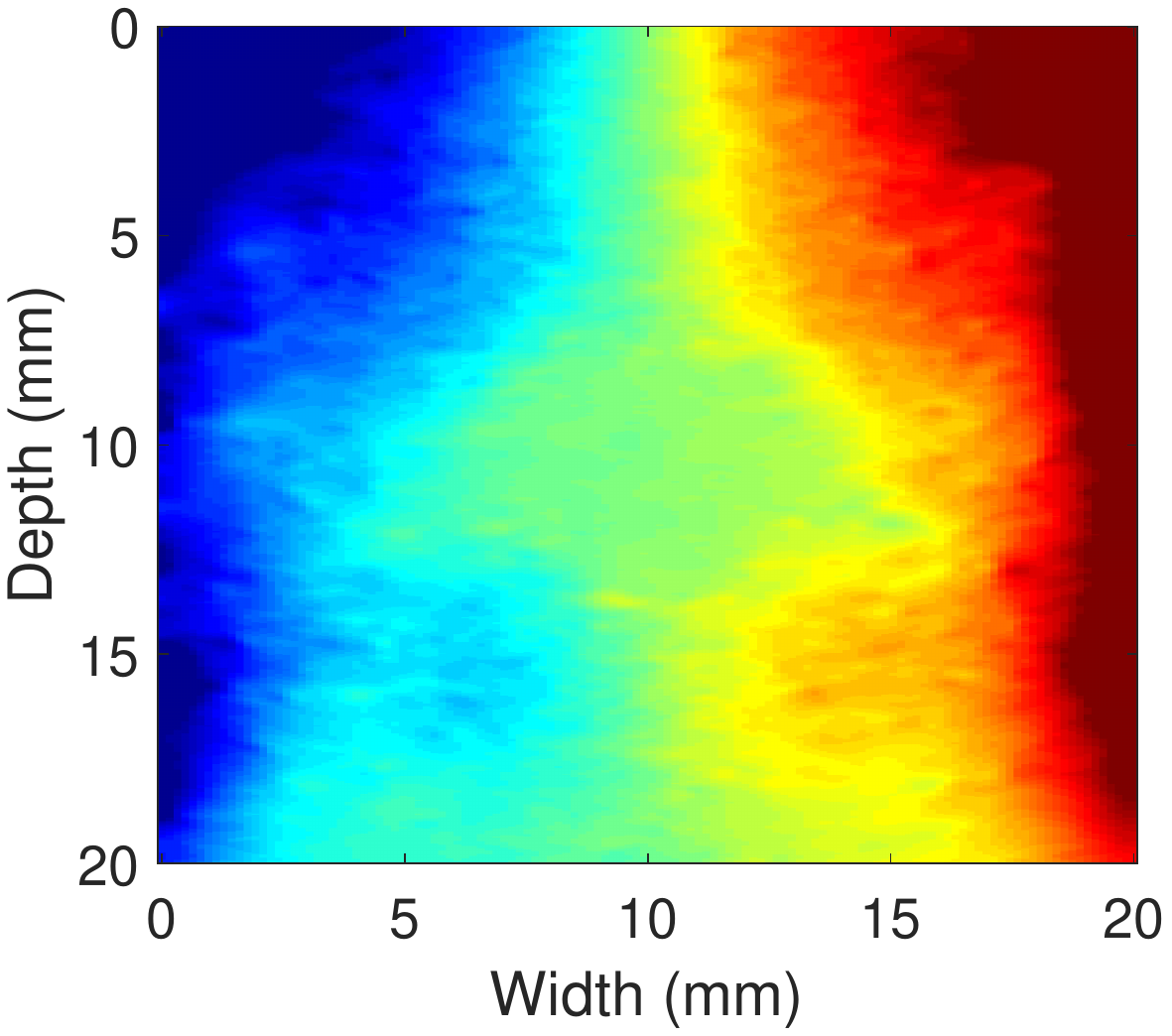}}}
	\subfloat[Inter. OVERWIND]{{\includegraphics[width=4cm]{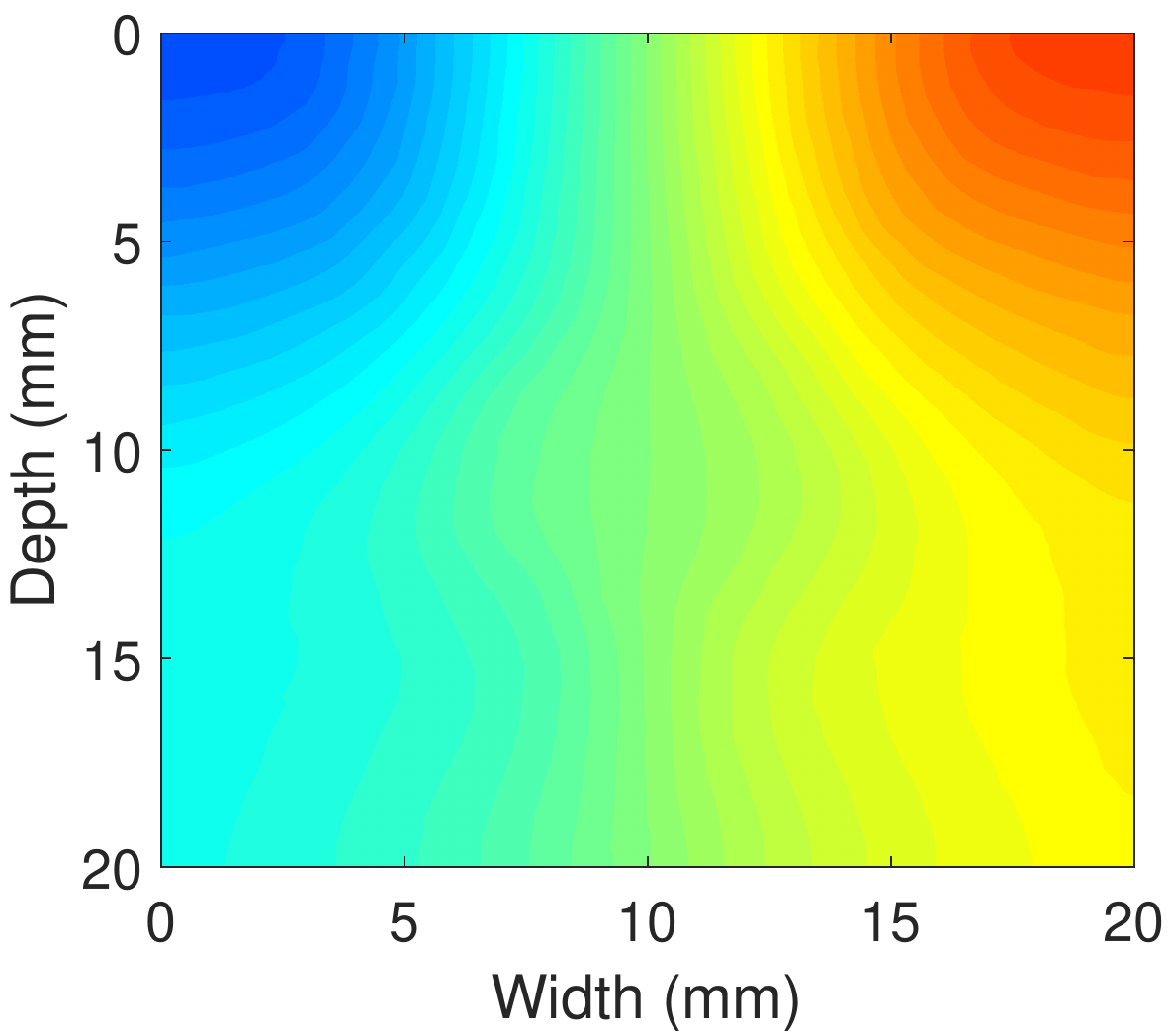}}}
	\subfloat[HF OVERWIND]{{\includegraphics[width=4cm]{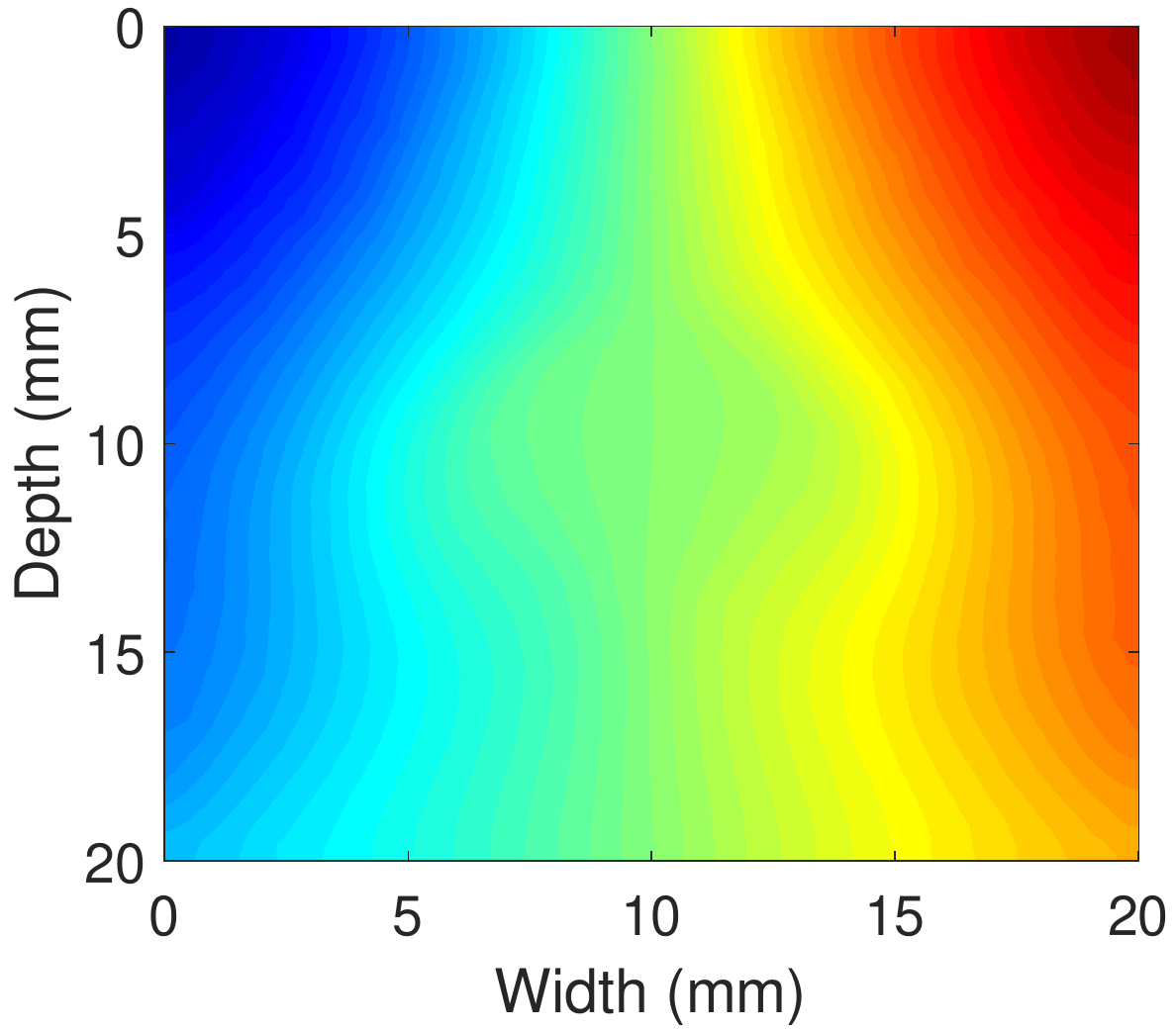}}}
	\subfloat{{\includegraphics[width=0.75cm]{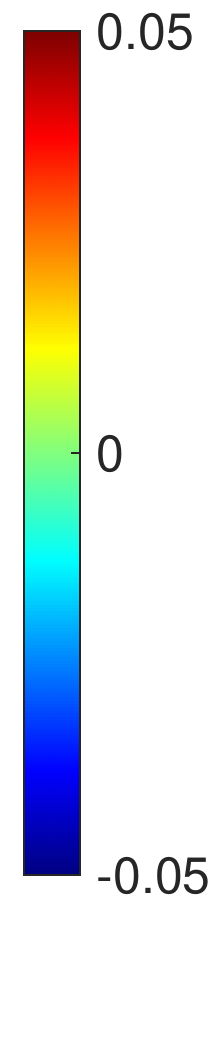}}}

	\subfloat[Ground truth]{{\includegraphics[width=4cm]{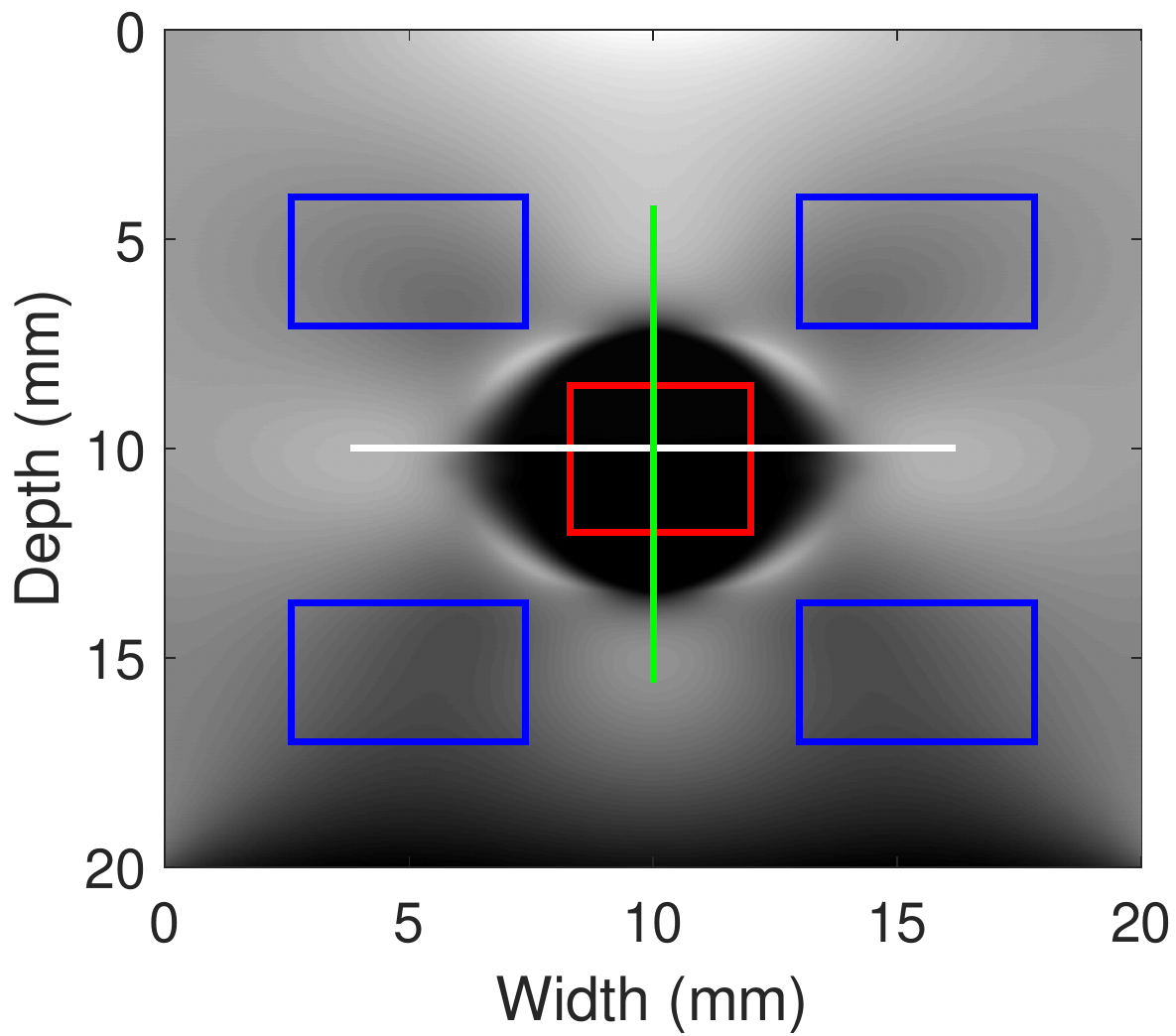}}}
	\subfloat[LF OVERWIND]{{\includegraphics[width=4cm]{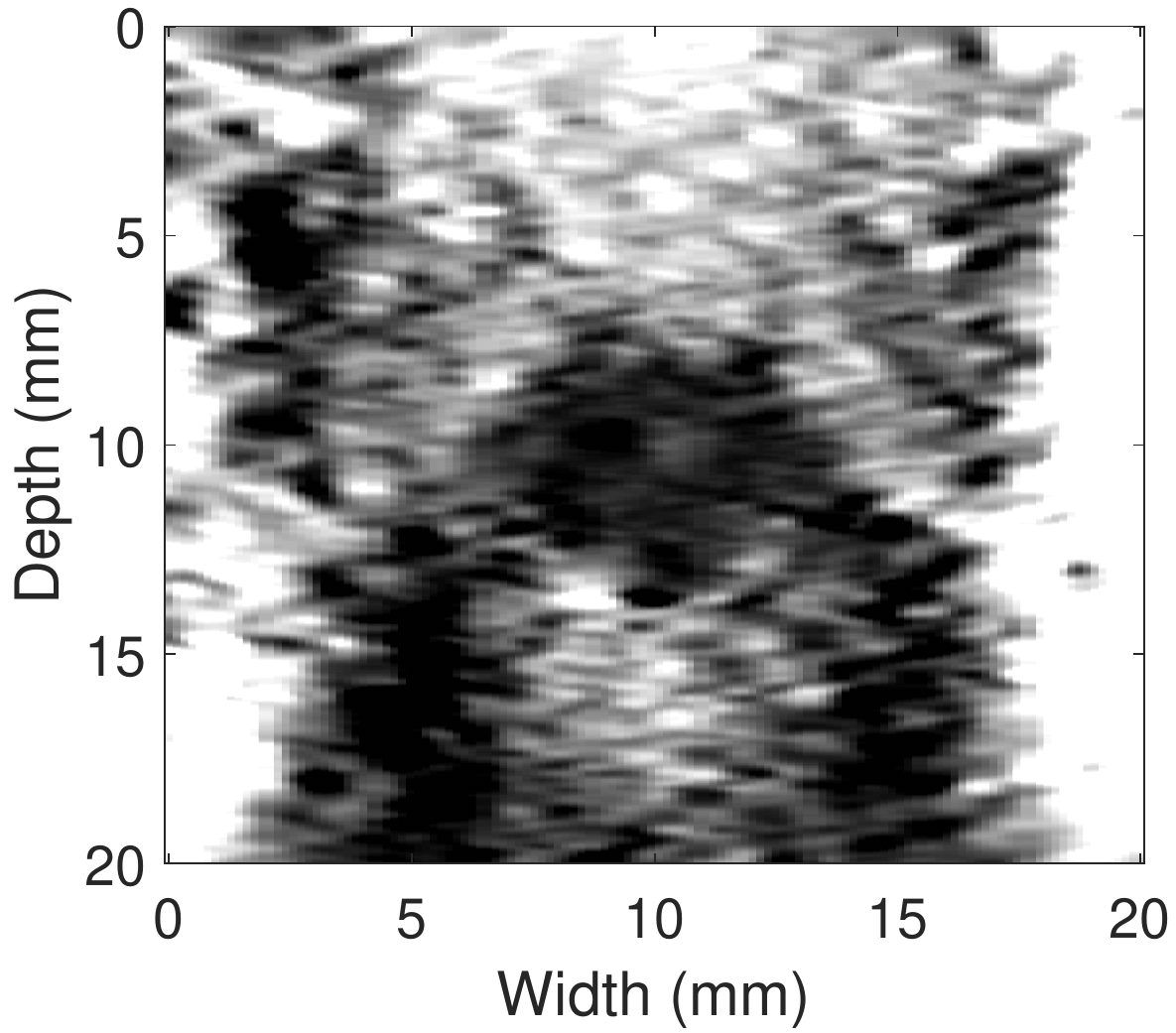}}}
	\subfloat[Inter. OVERWIND]{{\includegraphics[width=4cm]{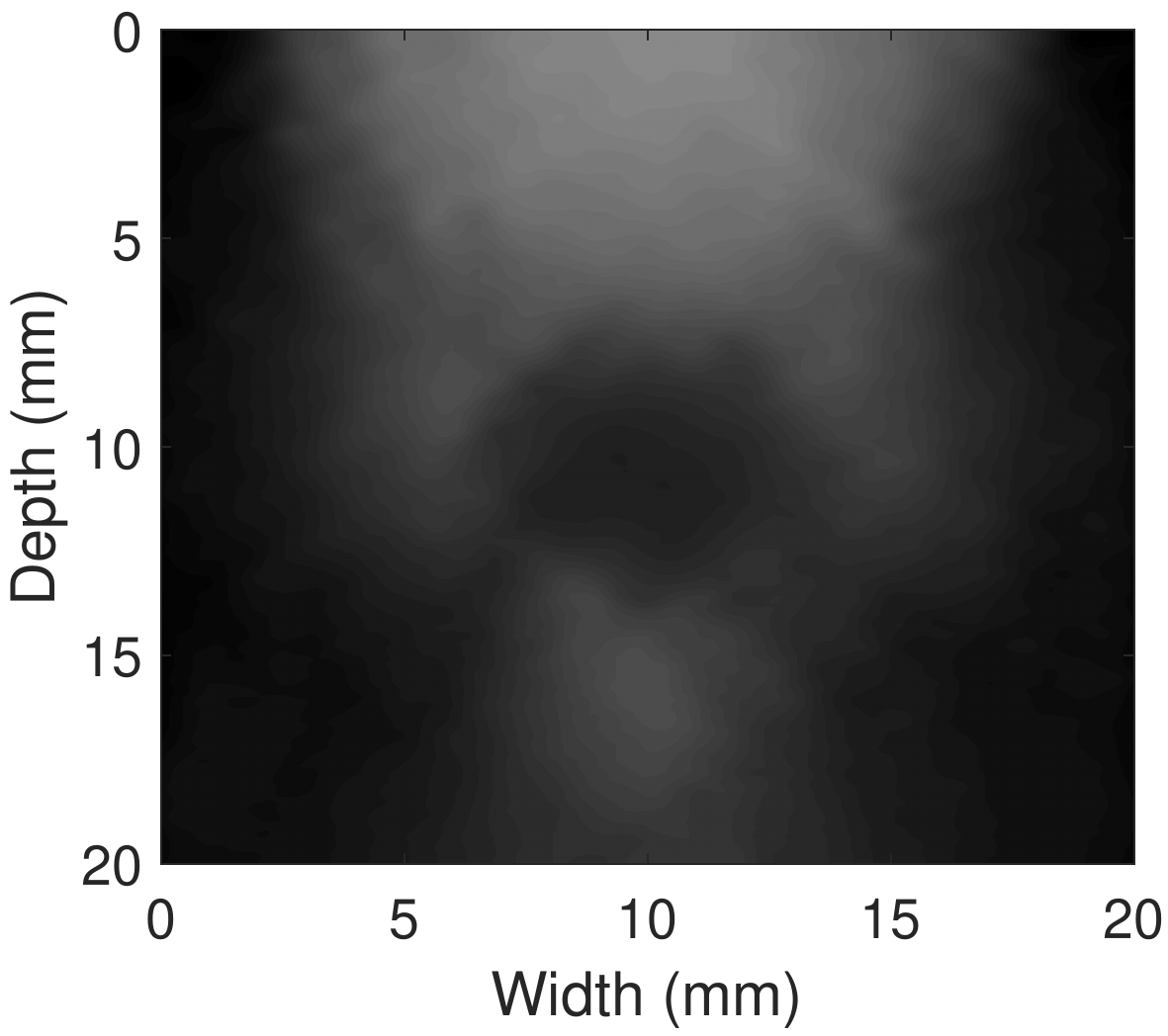}}}
	\subfloat[HF OVERWIND]{{\includegraphics[width=4cm]{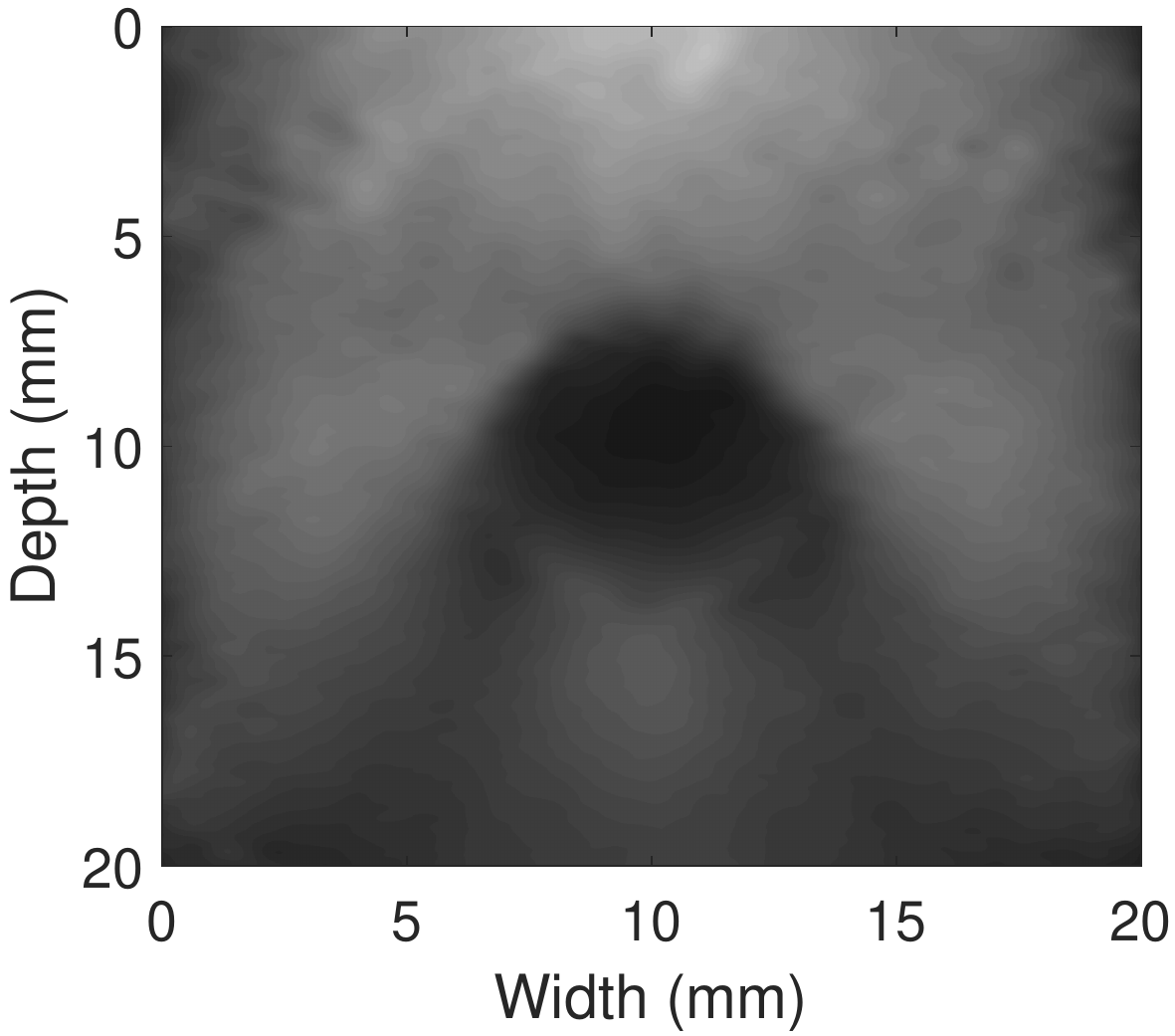}}}
	\subfloat{{\includegraphics[width=0.8cm]{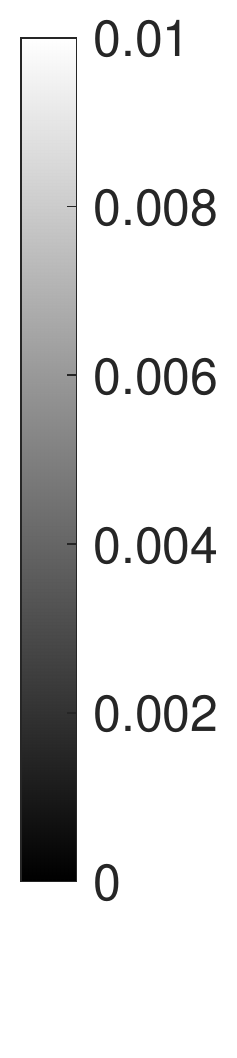}}}	
	\caption{ Estimated lateral displacement with LF OVERWIND, Inter. OVERWIND and HF OVERWIND for simulation data are shown in (b)-(d), respectively.} The second row shows the corresponding strains. The red and blue rectangles in (f) are considered as target and background areas for CNR calculation. The horizontal white and green vertical lines are also used for plotting the edge spread function of Fig.~\ref{esf_lateral}  
	\label{sim_virtual_lateral}
\end{figure*} 
\subsection{Line by Line Imaging}
In this imaging technique, a group of elements (i.e. transmission aperture) transmits the beam to increase the penetration of signal to image deeper areas. The transmitted beam focuses at a single point during each transmission. The data received at different channels is processed to generate one line of the US image. Fig. \ref{imaging}-(b) shows the pattern of transmission by dashed red lines, while the area that is imaged in each transmission is highlighted by gray. In this technique, the lateral resolution at the focal depth
is high and close to the resolution of the SA imaging. Further  away from this point, the resolution decreases. Moreover, the number of lines is limited to the number of elements (without interpolation).   

\subsection{VSSA: Virtual Source Synthetic Aperture}
This imaging technique benefits from both SA and line by line imaging and has the ability of increasing sampling frequency and resolution in lateral direction in all imaged area while penetrating to deep fields which are essential for ultrasound elastography. Similar to line by line imaging, a group of elements transmits the beam by focusing at a single point \cite{frazier1998synthetic,passmann1996100}. Since the  beam at the focal point is very narrow, we can assume the focal point as a virtual source that transmits the beam similar to SA as shown in Fig. \ref{imaging}-(c). Then each receiver focuses the received data at any point inside the aperture  according to  the following expression \cite{bottenus2013synthetic}
\begin{equation}
t_p({fj})=\dfrac{z_f\pm \sqrt{(x_p-x_f)^2+(z_f-z_p)^2}+\sqrt{(x_j-x_p)^2+z_p^2}}{c}
\label{virtual_focus}
\end{equation}
where $c$ is the speed of sound in soft tissue, $x_p$, $x_f$, $z_p$ and $z_f$  are the positions of the focal point $f$ and $p$ is the location at which the beam is focused. The $\pm$ term in Eq. \ref{virtual_focus} divides the imaging area into regions above and below the virtual source.
Similar to SA, summation of focused data by different receivers generates a low-resolution image and by switching the transmissions new focal points will be the virtual sources and in each step a new low resolution image will be generated. Similar to SA, adding up these images will end up with a high-resolution image as
\begin{equation}
y_p=\sum\limits_{f=1}^e \sum\limits_{j=1}^e t_p({fj})
\end{equation}

Similar to line by line imaging, the number of A-lines is usually equal to number of elements in VSSA, and interpolation is the most commonly used technique for increasing the sampling frequency in the lateral direction for elastography purposes.  However, for the VSSA the received data can be focused at any point inside the aperture (highlighted by gray in Fig. \ref{imaging}-(c)). To increase the sampling frequency, we consider a grid consisting of nodes with same spatial distance of nodes in axial and lateral direction as $p_d=\frac{c}{2f_s}$. Each receiver element generates an image on that grid as shown in Fig. \ref{imaging}-(c). Therefore, we can increase number of data and resolution in lateral direction without interpolation.

\subsection{Data Acquisition and comparison metrics}
In this section, the data that are utilized in different experiments of the paper are described and then results of HF OVERWIND are compared with Inter. OVERWIND  and also LF OVERWIND wherein number of data in lateral direction is equal to number of piezo-electrics. In the Results Section, the CNR metric is used to provide a quantitative value for assessing the proposed method \cite{varghese1998analysis} 
\begin{equation}
\textrm{CNR}=20\ \textrm{log}_{10}{\Big(\dfrac{2(\bar{s}_b-\bar{s}_t)^2}{\sigma_b^2+\sigma_t^2}\Big)}, 
\label{eq1}
\end{equation} 
where $\bar{s}_t$ and $\bar{s}_b$ are the spatial strain averages of the target and
background, $\sigma_t^2$ and $\sigma_b^2$ are the spatial strain variances of the
target and background, respectively \cite{ophir1999elastography}. 
For the simulation results where we know the ground truth, we use Root Mean Square Error (RMSE), Mean of estimation Error (ME) and Variance of estimation Error (VE) as other metrics according to   
\begin{equation}
\begin{array}{c}
\textrm{RMSE\%}=100*\dfrac{\sqrt{m\times n\times \sum\limits_{i=1}^m\sum\limits_{j=1}^n\big(S_e(i,j)-S_g(i,j)\big)^2}}{\sum\limits_{i=1}^m\sum\limits_{j=1}^nS_g(i,j)},\\
\textrm{ME}=\dfrac{\sum\limits_{i=1}^m\sum\limits_{j=1}^nS_e(i,j)-S_g(i,j)}{m\times n}\\
\textrm{VE}=\dfrac{\sum\limits_{i=1}^m\sum\limits_{j=1}^n\big(S_e(i,j)-S_g(i,j)\big)^2}{m\times n}-ME^2
\end{array}
\label{eq2}
\end{equation} 
where $m$ and $n$ are size of estimated strains. $S_e$ and $S_g$ are estimated and ground truth strains, respectively.
\begin{figure}
	\centering
	\subfloat[horizontal]{{\includegraphics[width=4.2cm]{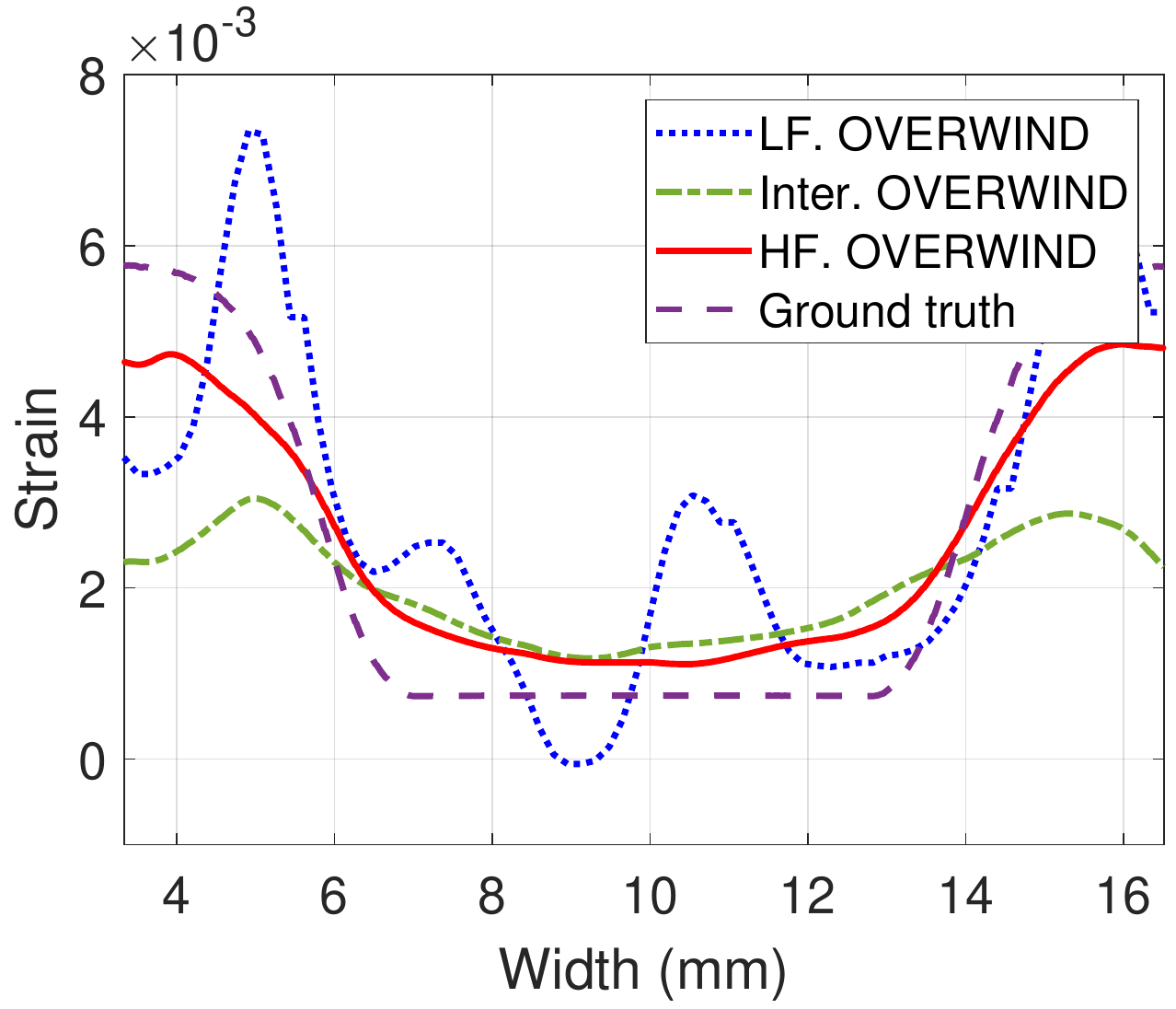}}}
	\subfloat[vertical]{{\includegraphics[width=4.2cm]{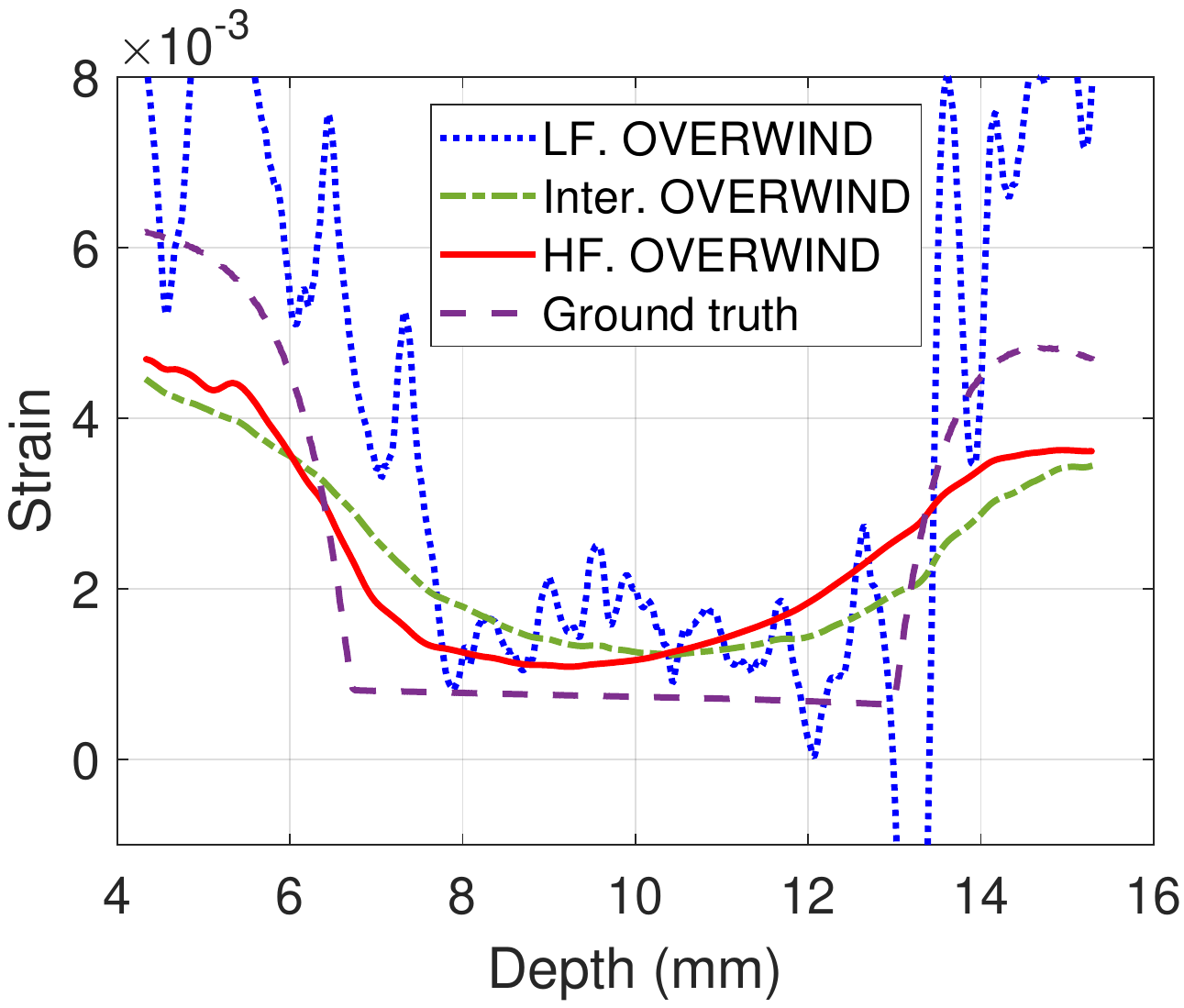}}}	
	\caption{Edge spread function of the lateral strain in horizontal (a) and vertical line (b).}
	\label{esf_lateral}
\end{figure}
\begin{figure*}
	\centering
	\subfloat[Ground truth]{{\includegraphics[width=4cm]{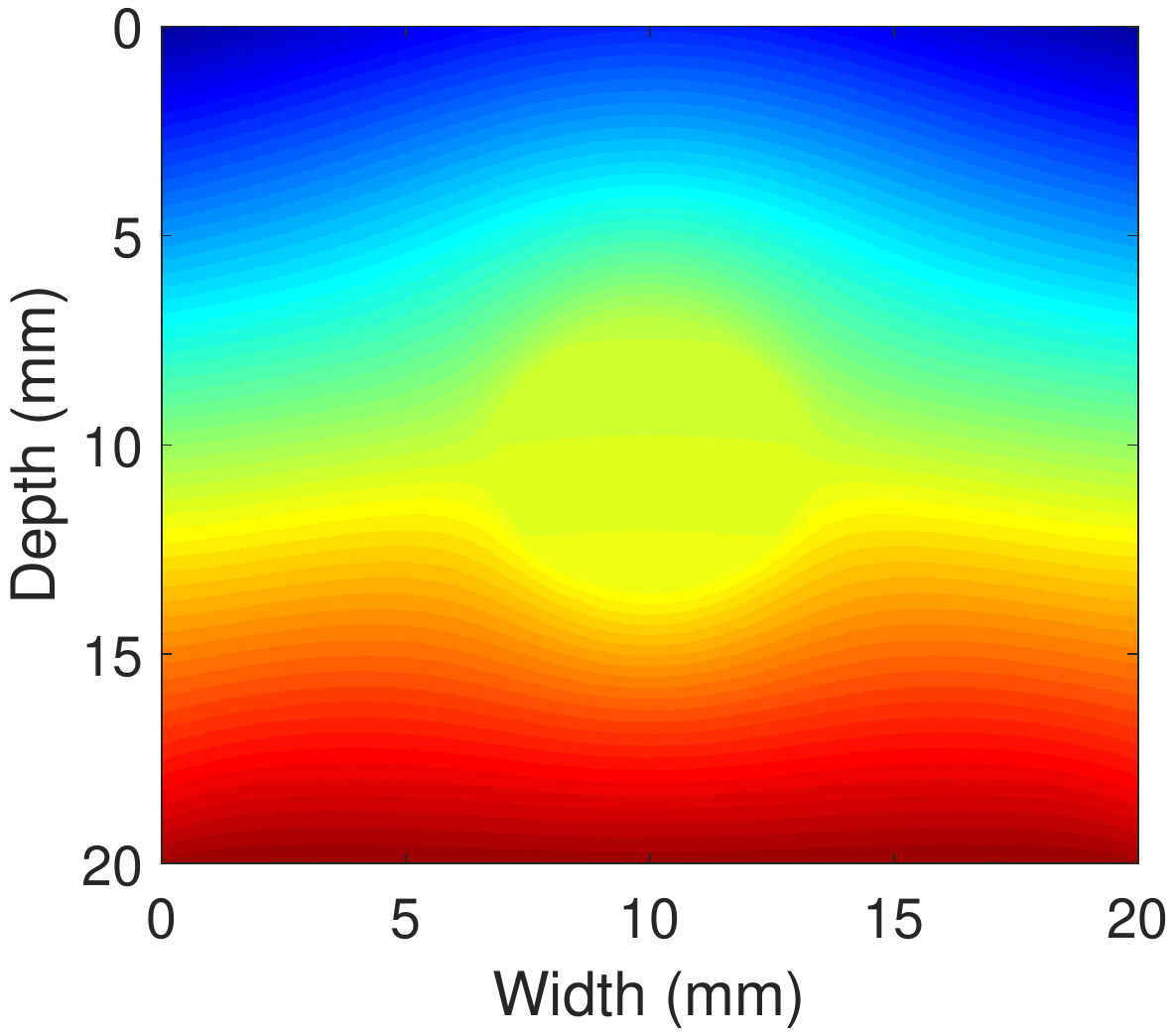}}}
	\subfloat[LF OVERWIND]{{\includegraphics[width=4cm]{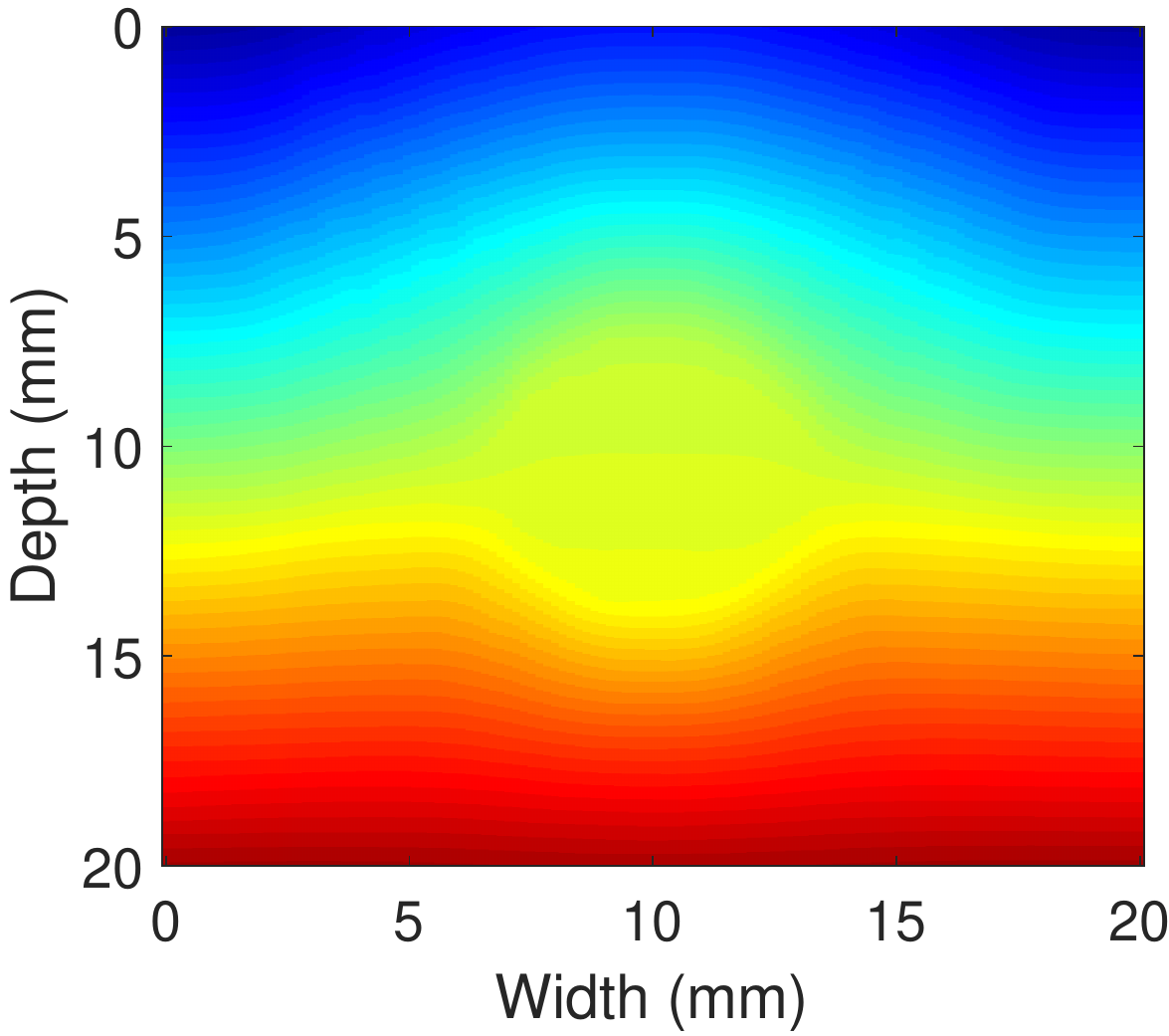}}}
	\subfloat[Inter. OVERWIND]{{\includegraphics[width=4cm]{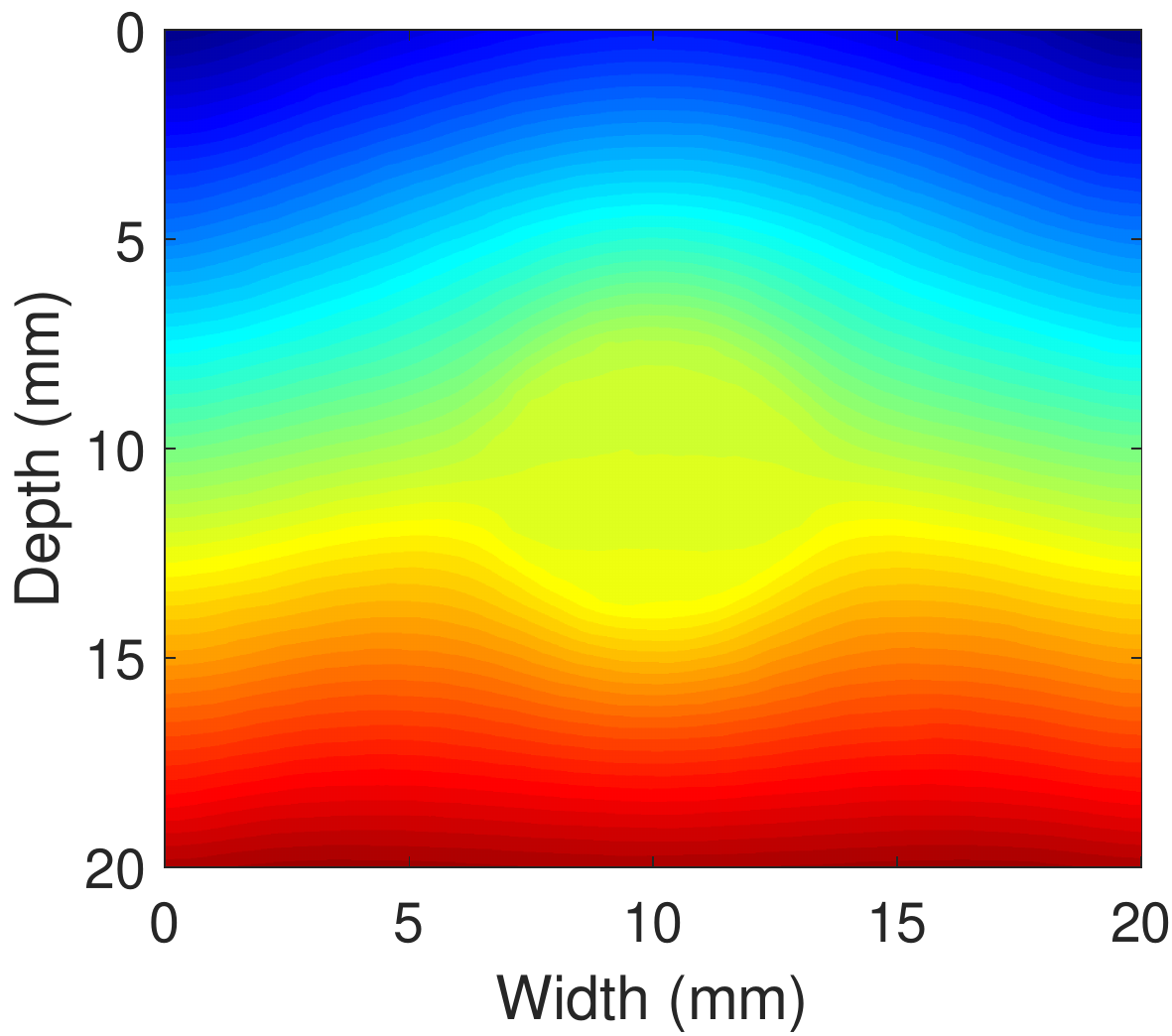}}}
	\subfloat[HF OVERWIND]{{\includegraphics[width=4cm]{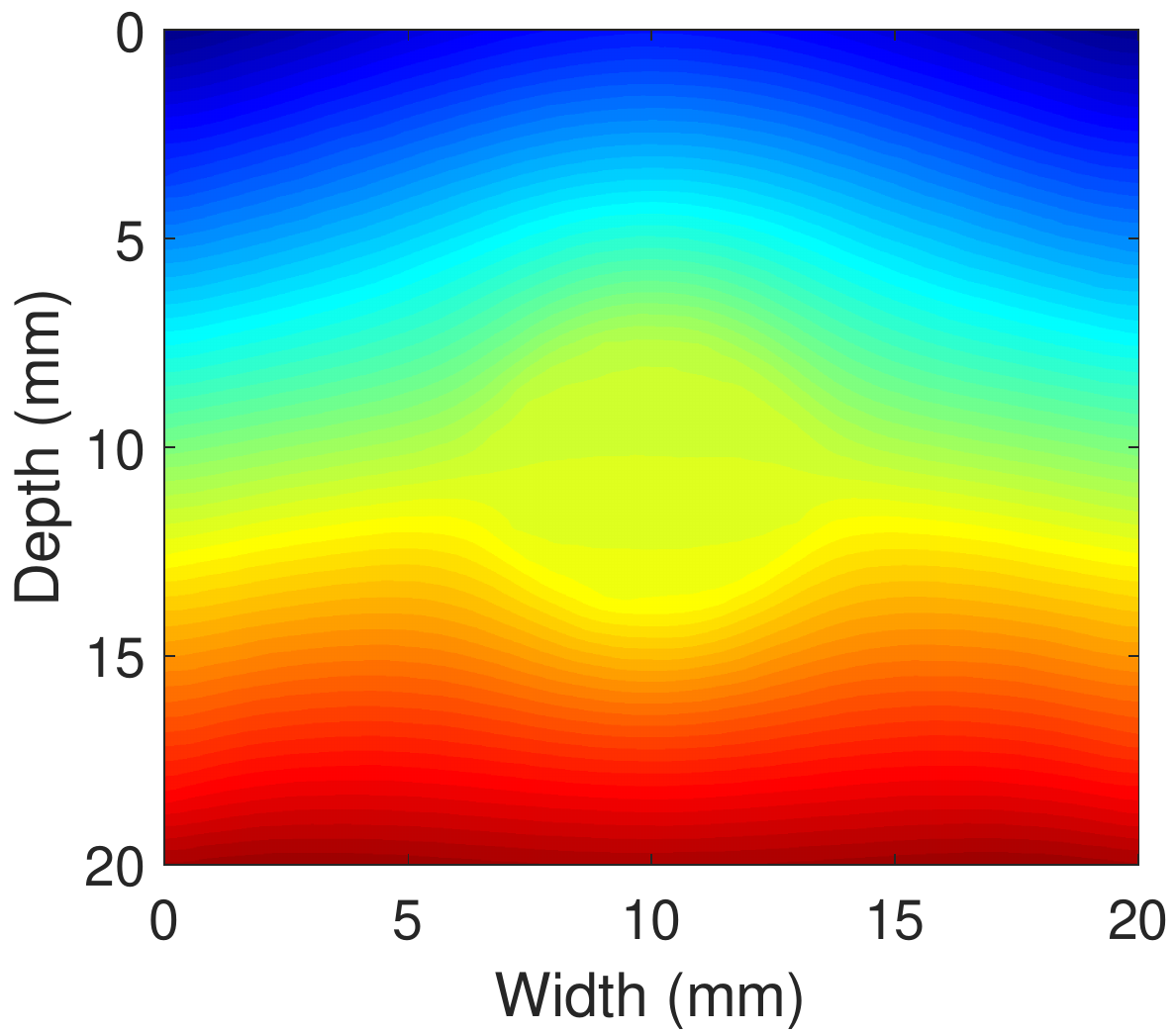}}}
	\subfloat{{\includegraphics[width=0.68cm]{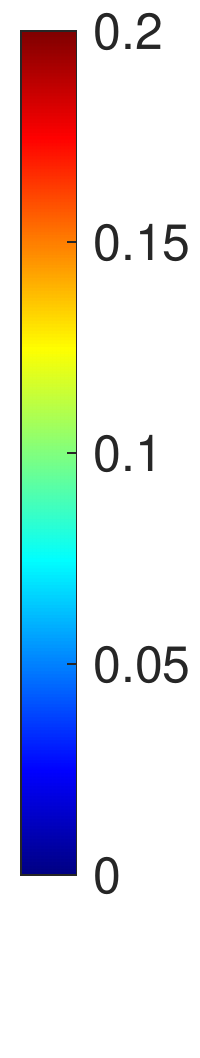}}}
	
	\subfloat[Ground truth]{{\includegraphics[width=4cm]{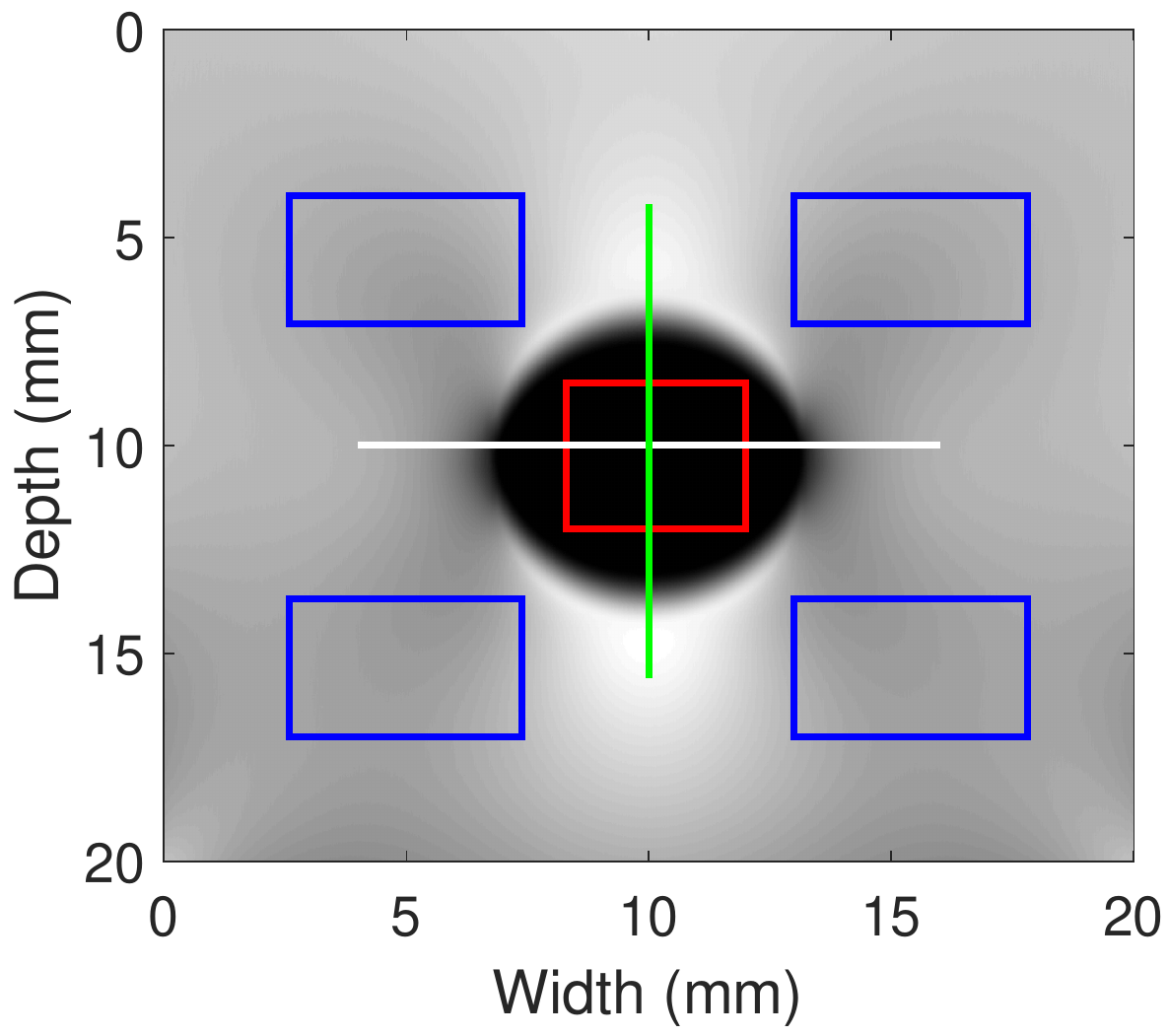}}}
	\subfloat[LF OVERWIND]{{\includegraphics[width=4cm]{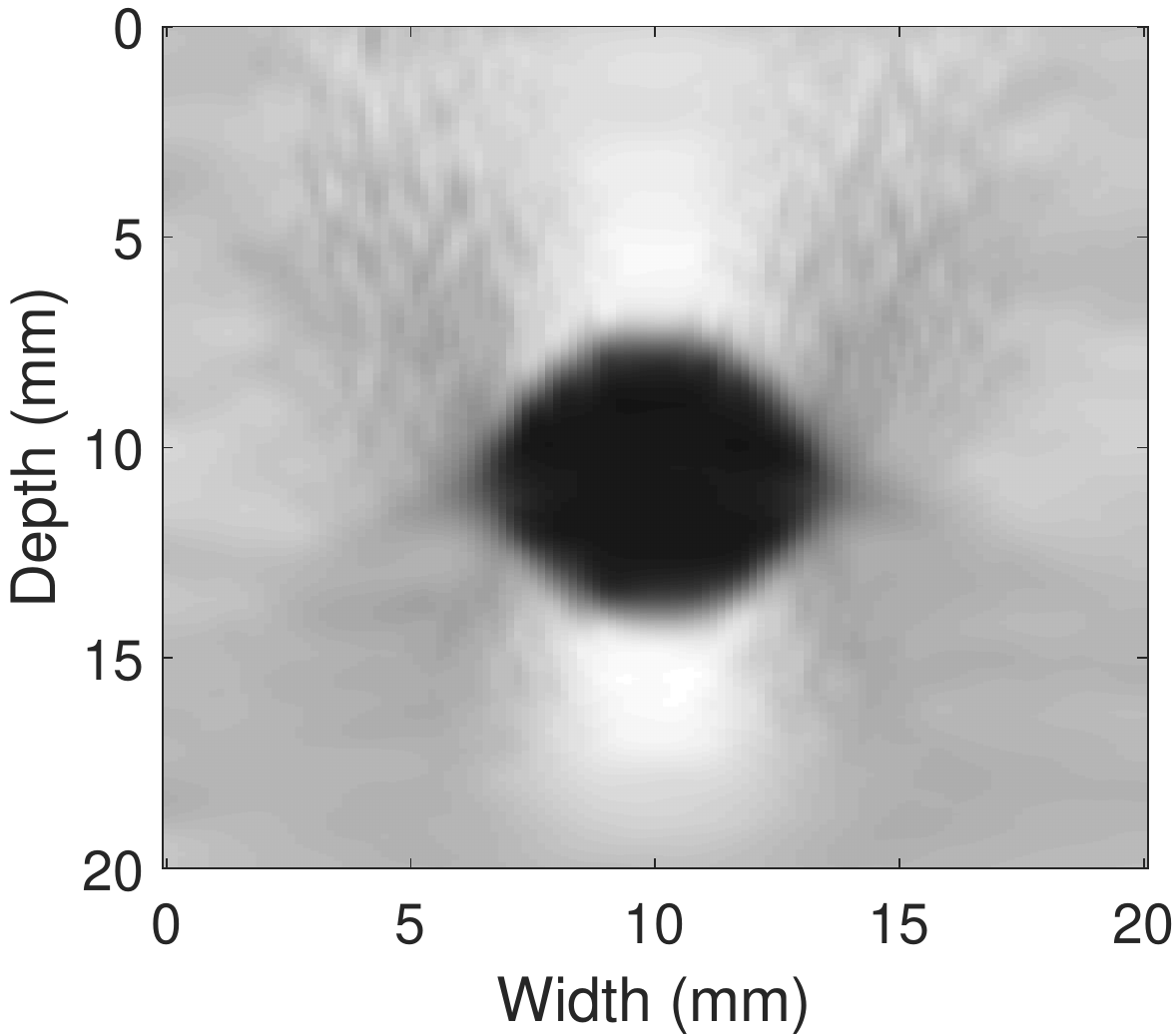}}}
	\subfloat[Inter. OVERWIND]{{\includegraphics[width=4cm]{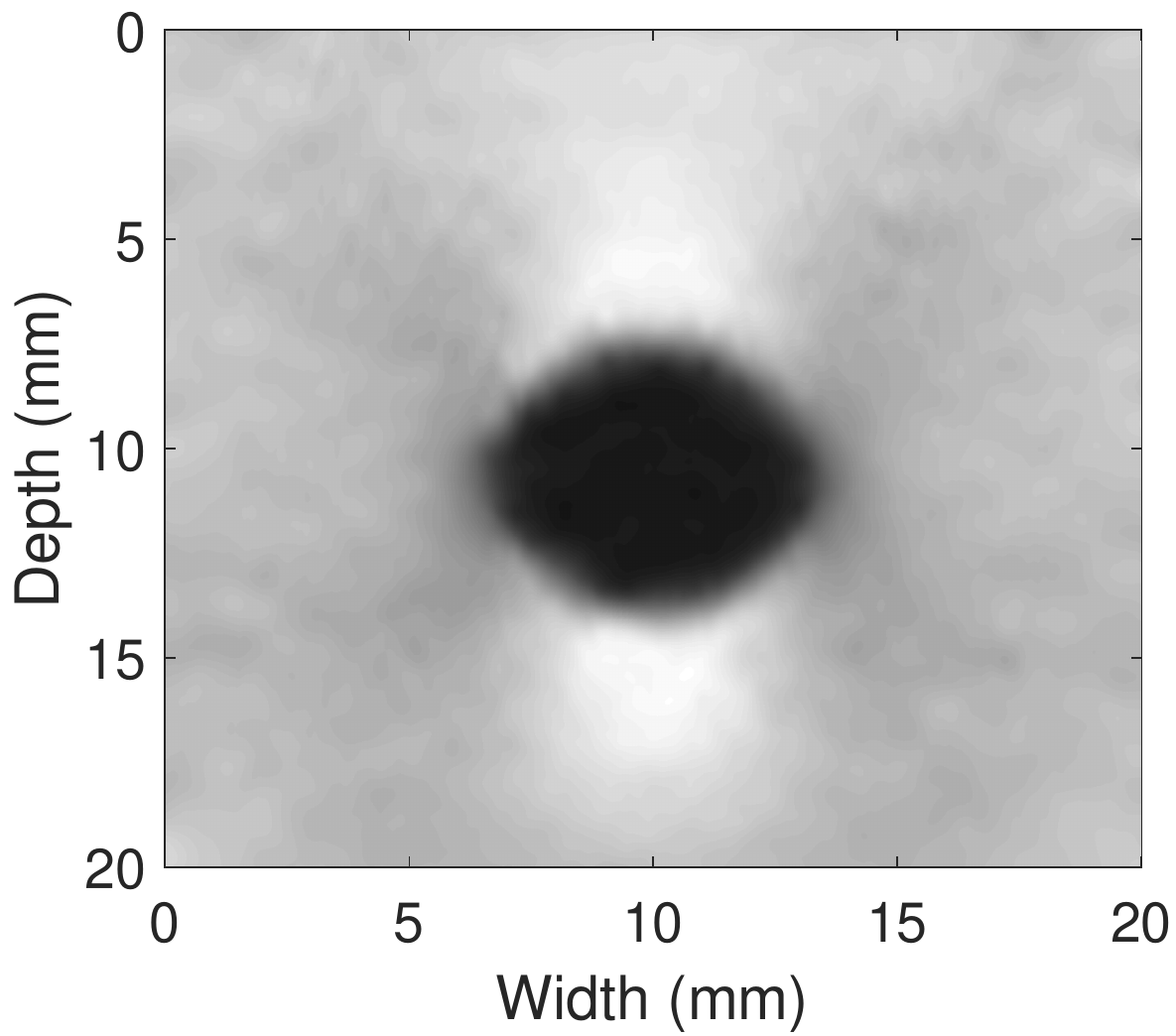}}}
	\subfloat[HF OVERWIND]{{\includegraphics[width=4cm]{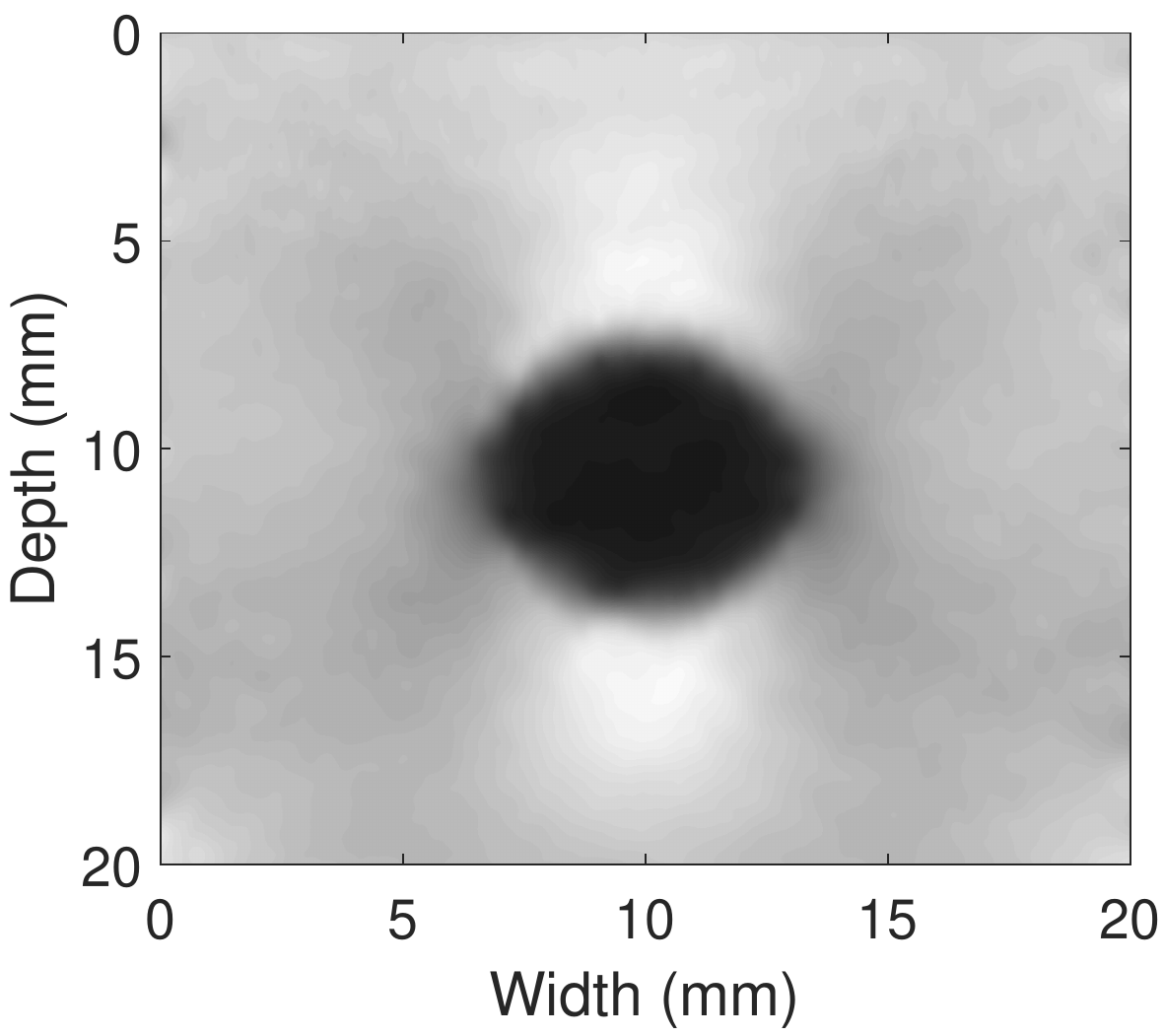}}}
	\subfloat[]{{\includegraphics[width=0.86cm]{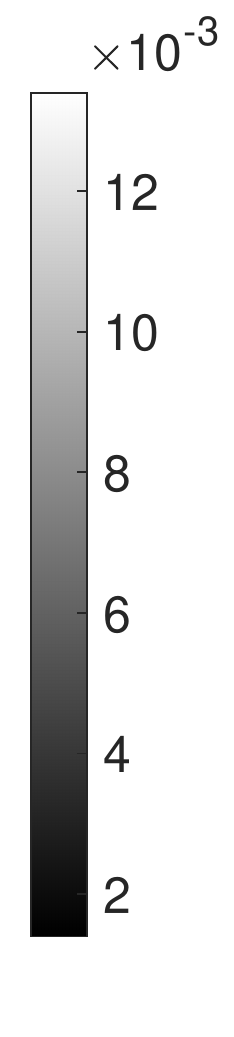}}}	
	\caption{Estimated axial displacement with LF OVERWIND, Inter. OVERWIND and HF OVERWIND for simulation data are shown in (b)-(d), respectively. The second row shows the corresponding strains.  The red and blue rectangles in (f) are considered as target and background areas for CNR calculation. The horizontal white and green vertical lines are also used for plotting the edge spread function of Fig.~\ref{esf_axial}  }
	\label{sim_virtual_axial}
\end{figure*}
For estimating the axial strain, the displacement field should be differentiated.  To reduce the impact of noise during differentiating, it is common to use Least Square Estimation (LSQ) for strain estimating. For estimating the strain at each sample, a few neighboring samples in a window of size $\rho$ are considered and a line is fitted to their displacements. The tangent of the line is considered as the strain for the  middle sample. Considering more data points for least square makes the strain smooth at the cost of losing resolution. Throughout the paper, the size of LSQ window is $5\%$ of total data size.
\begin{figure}
	\centering
	\subfloat[horizontal]{{\includegraphics[width=4cm]{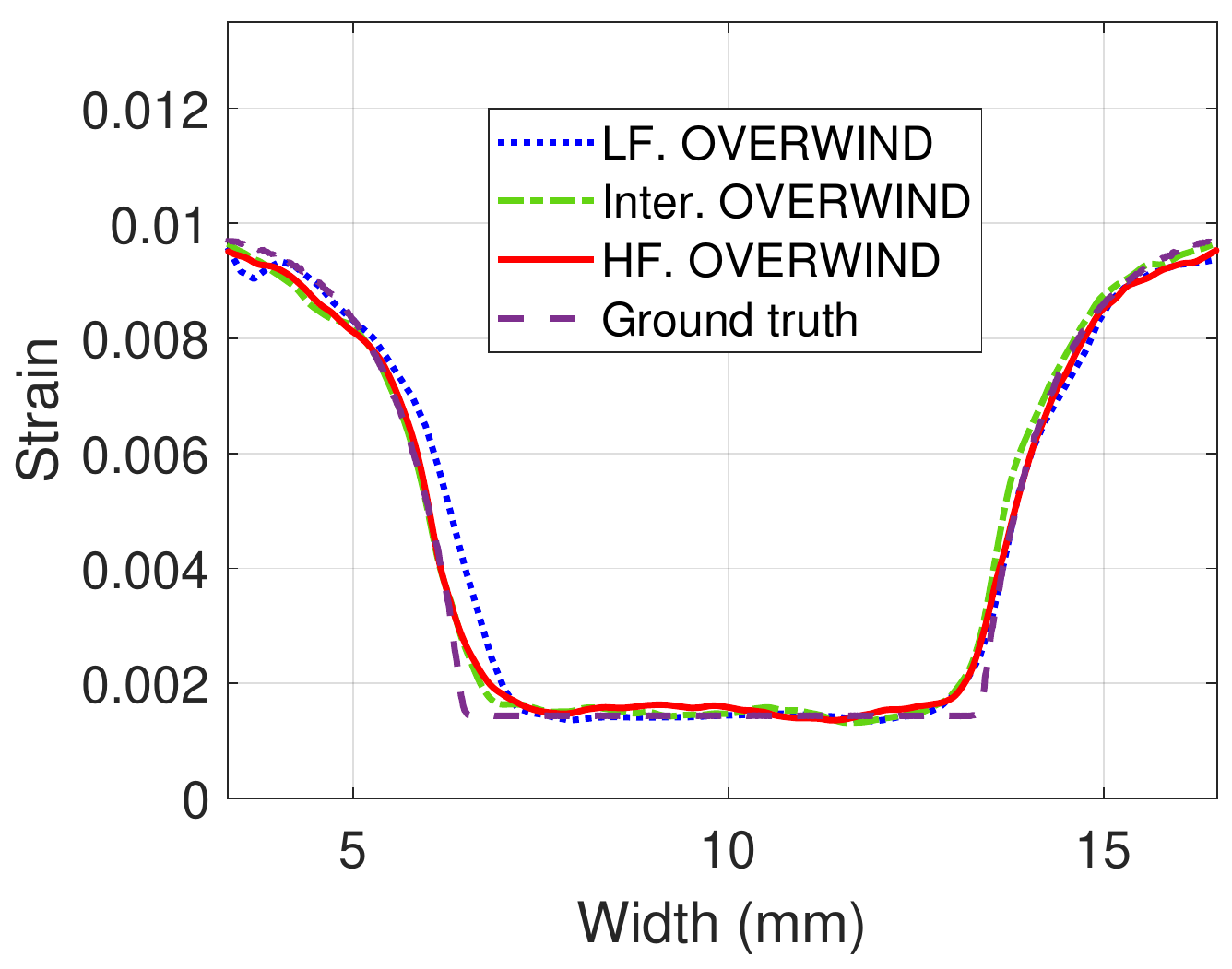}}}
	\subfloat[vertical]{{\includegraphics[width=4cm]{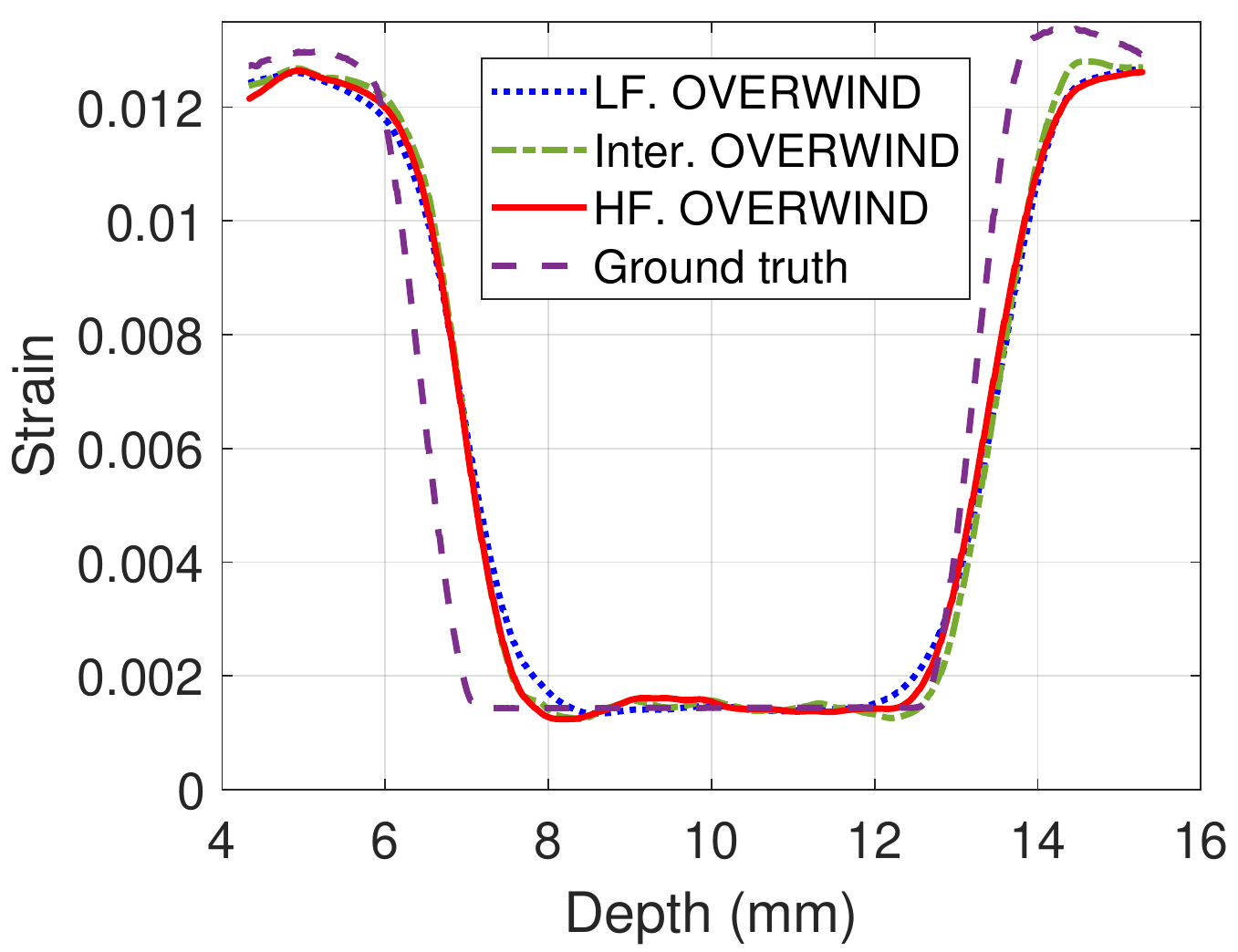}}}	
	\caption{Edge spread function of the axial strain in horizontal (a) and vertical line (b).}
	\label{esf_axial}
\end{figure}
\subsection*{Simulation Data}
A simulated  phantom is generated by utilizing the Field II ultrasound simulation software~\cite{fielii,139123} by randomly distributing 
	slightly more than $10$ scatterers per resolution cell
	to satisfy the Rayleigh scattering regime.
	
	The simulated phantom consists of a homogeneous region with a Young's modulus of $4\ kPa$ and one cylindrical inclusion with a Young's modulus of $40\ kPa$. For compressing the phantom and computing its ground truth displacement, Finite Element Method (FEM)-based deformations are computed using  the ABAQUS  software package (Johnston, RI, USA) with triangular mesh sizes of $0.05$ $\textrm{mm}^2$.
	The probe consists of 128 elements with pitch of $0.15$ mm. The center frequency is $7$ MHz, while the sampling rate is $100$ MHz. The lateral sampling frequency in HF OVERWIND is 19 times higher  than LF OVERWIND. Therefore, the data is interpolated by a factor of 19 and using a cubic spline method for Inter. OVERWIND, so that Inter. OVERWIND and HF OVERWIND have the same number of samples.
	
	 \begin{figure*}
		\centering
		\begin{tabular}{ccccc}
			\subfloat[B-MODE]{{\includegraphics[width=4cm]{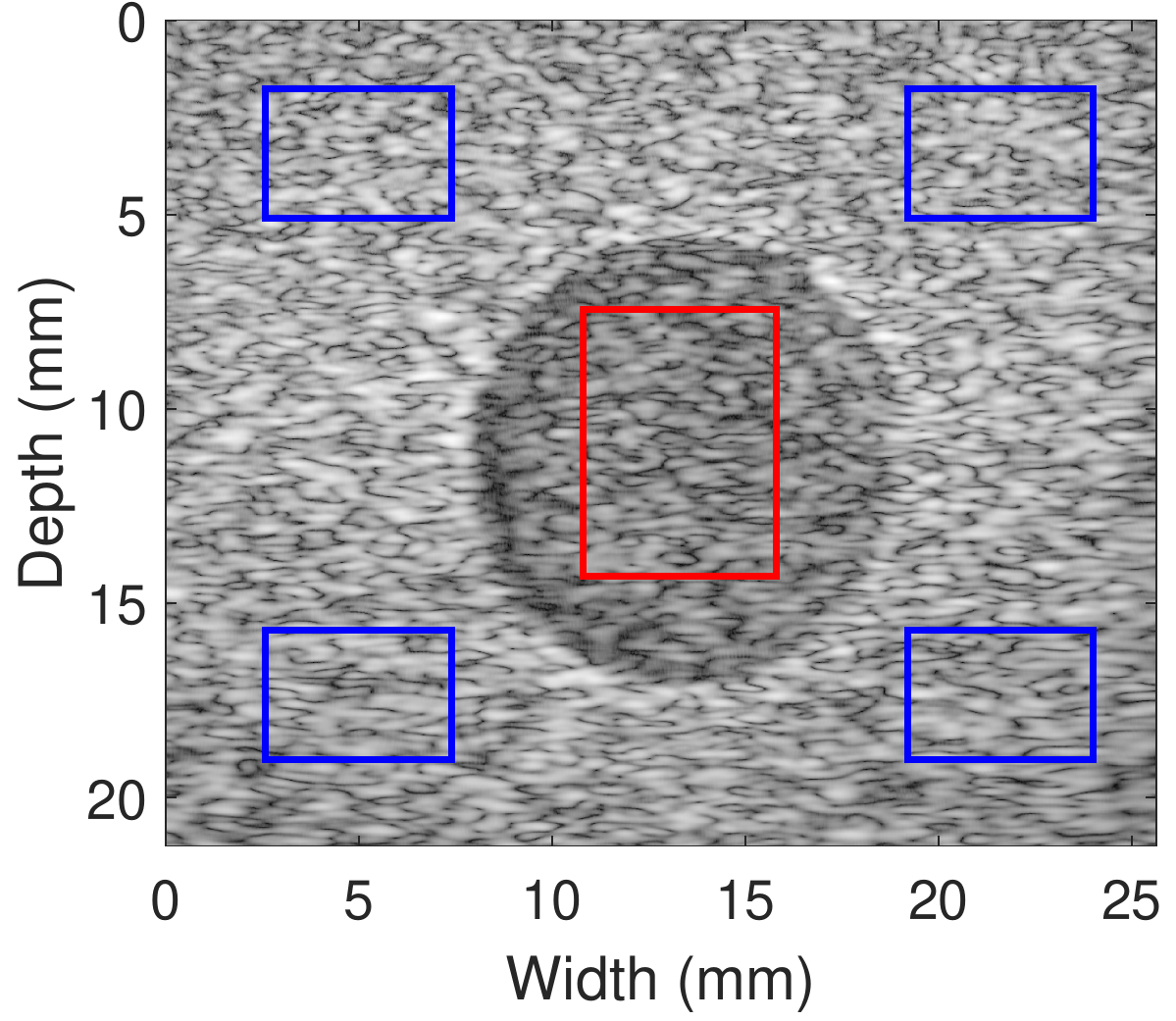}}}
		&\subfloat[LF OVERWIND]{{\includegraphics[width=4cm]{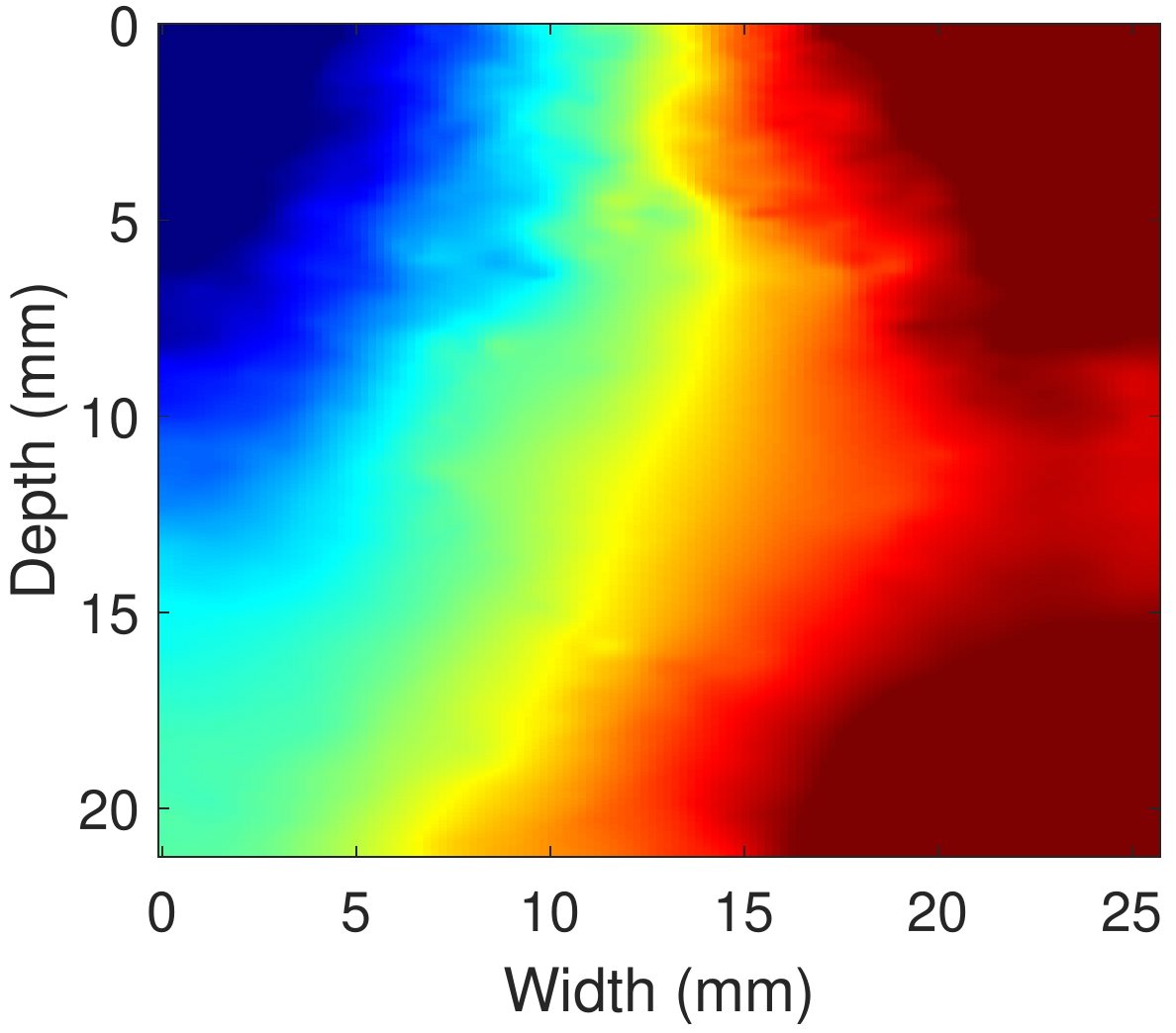}}}
	&\subfloat[Inter. OVERWIND]{{\includegraphics[width=4cm]{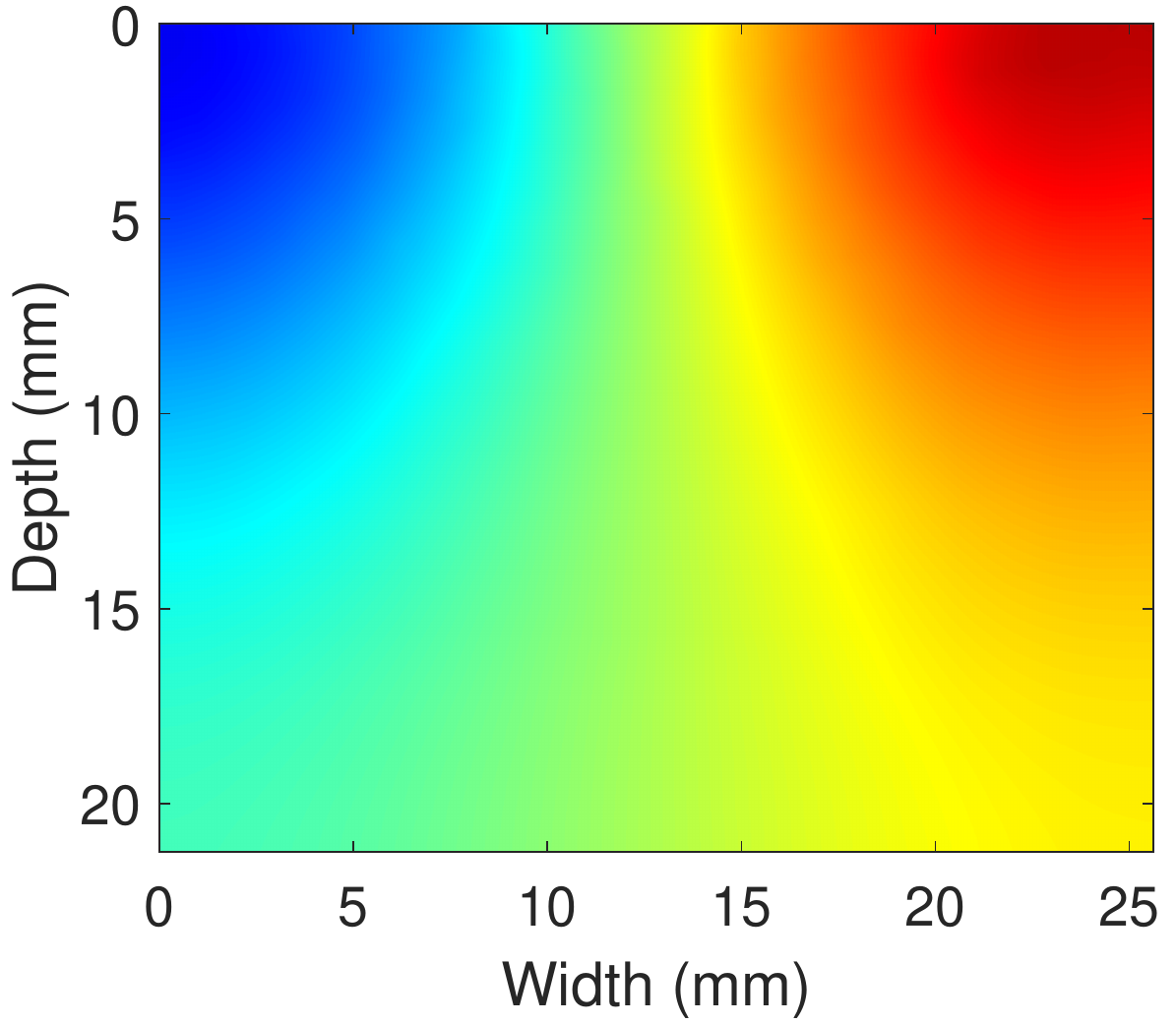}}}
	&\subfloat[HF OVERWIND]{{\includegraphics[width=4cm]{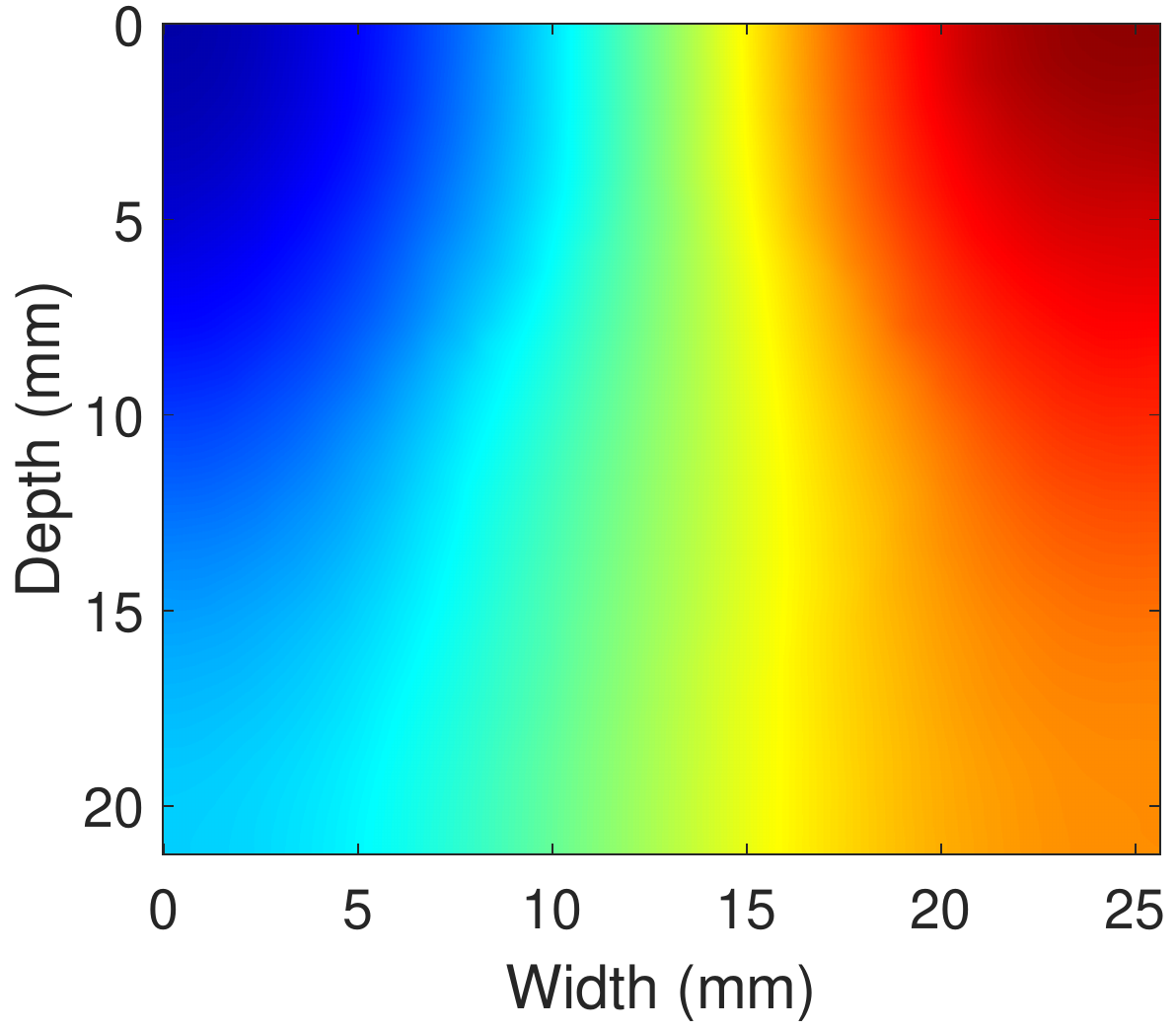}}}
		&\subfloat{{\includegraphics[width=0.64cm]{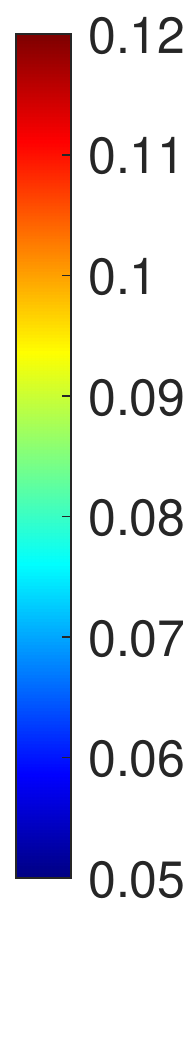}}}\\
		&\subfloat[LF OVERWIND]{{\includegraphics[width=4cm]{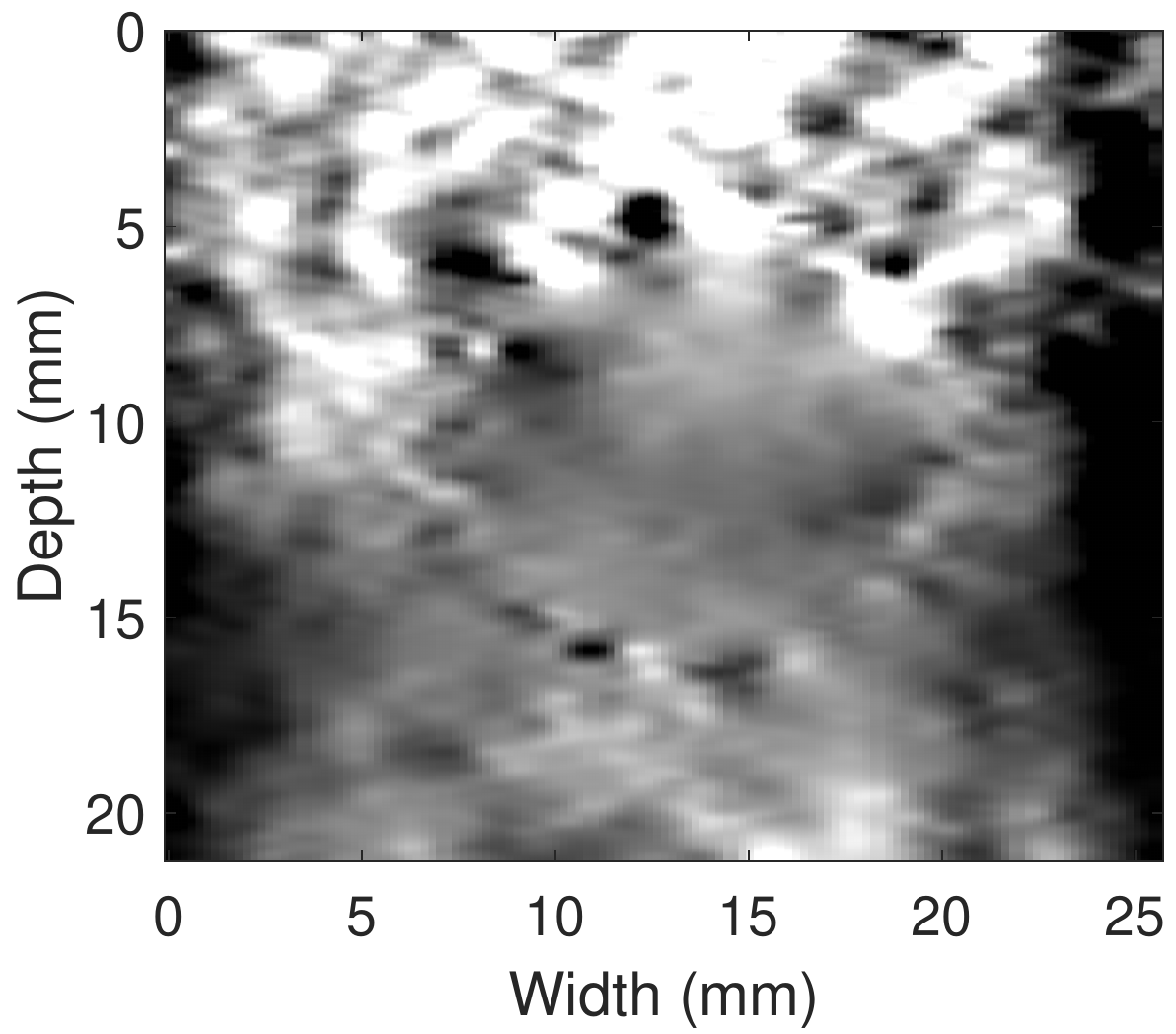}}}&
	\subfloat[Inter. OVERWIND]{{\includegraphics[width=4cm]{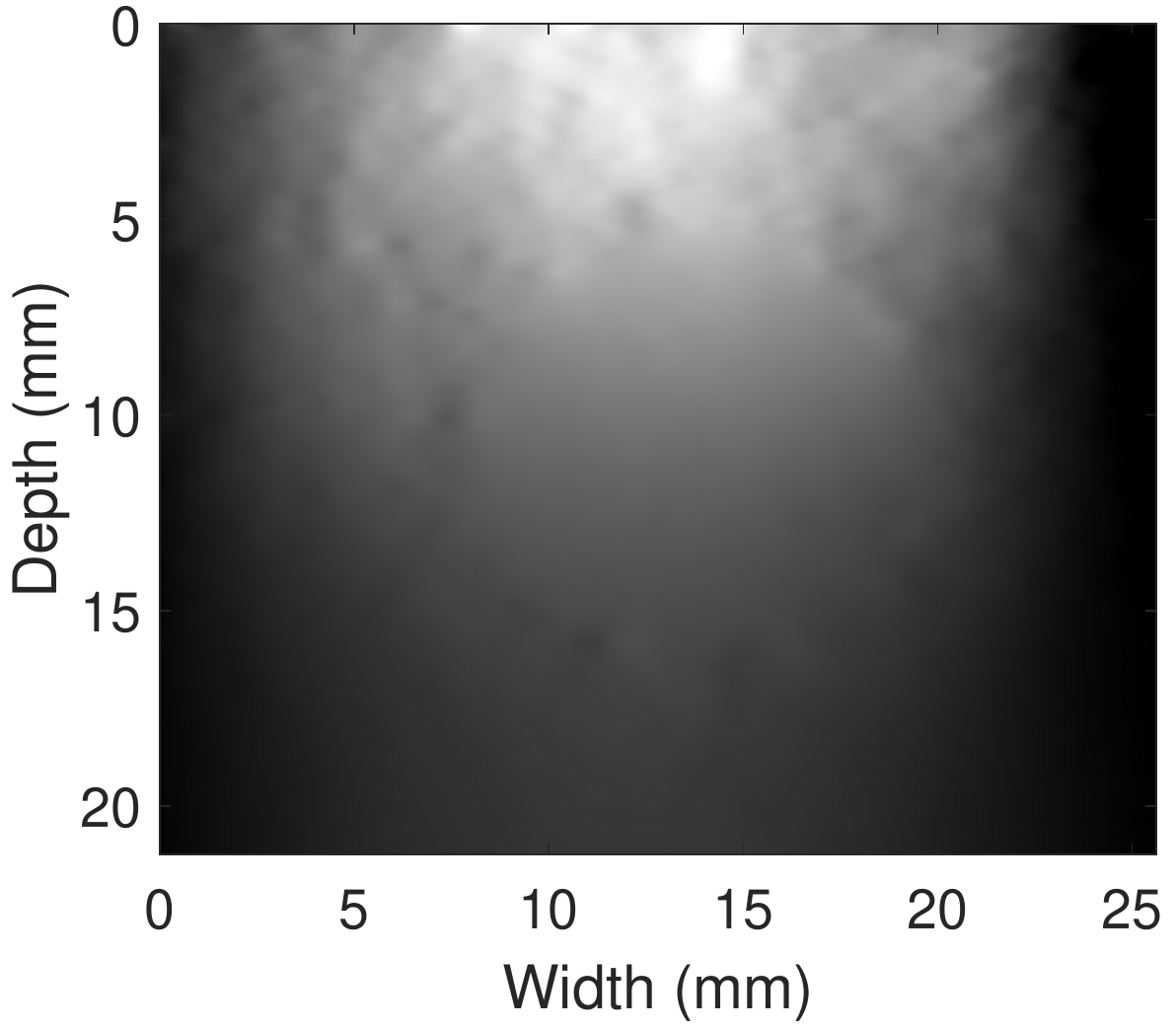}}}&
	\subfloat[HF OVERWIND]{{\includegraphics[width=4cm]{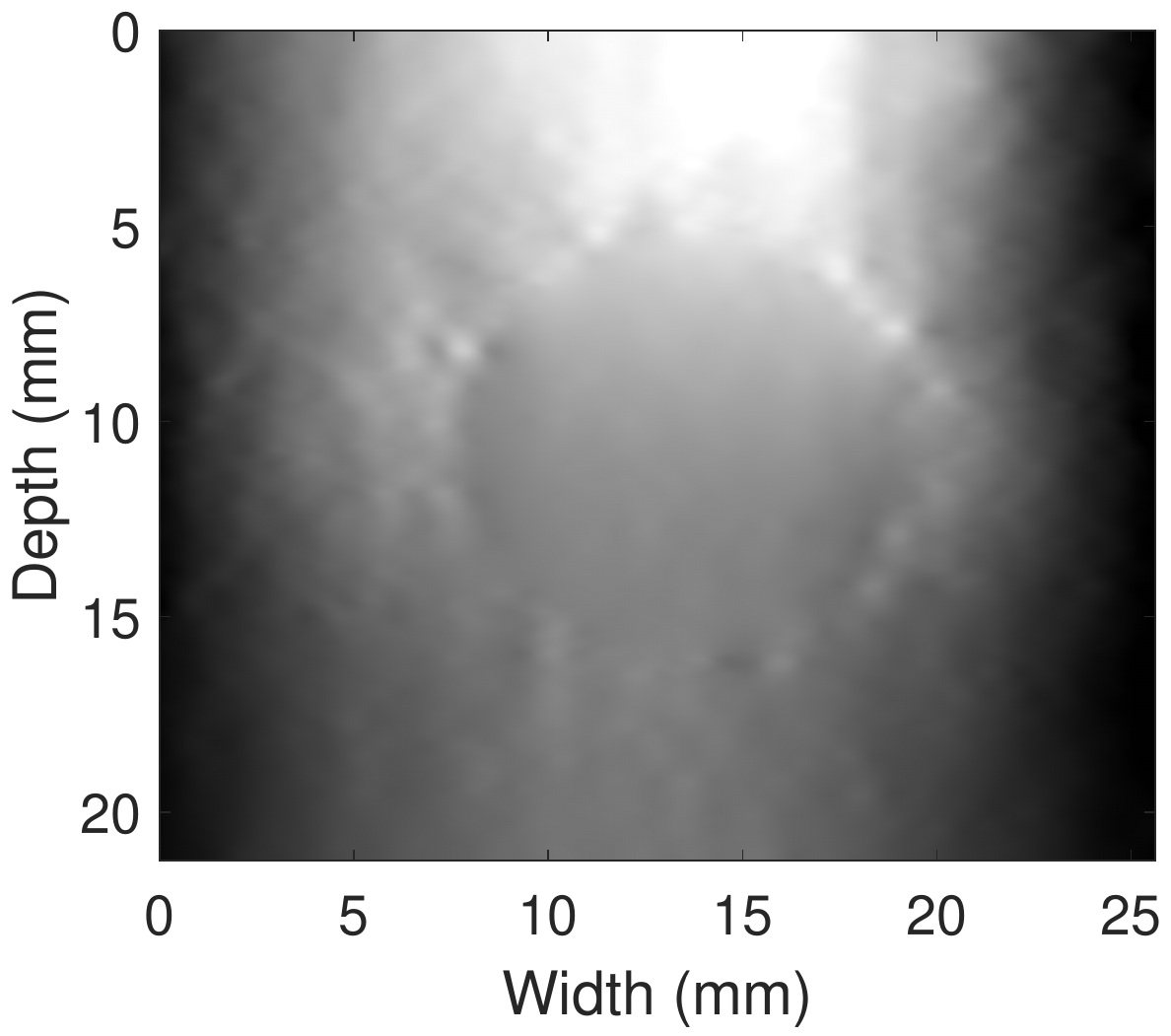}}}&
			\subfloat{{\includegraphics[width=0.74cm]{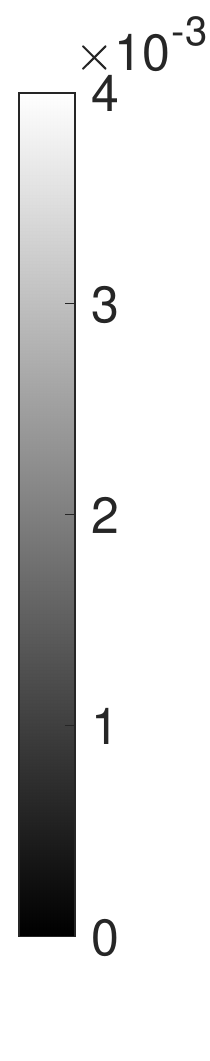}}}\\
		\end{tabular}
		\caption{Results on a tissue mimicking phantom. B-Mode image is shown in (a). Estimated lateral displacement with LF OVERWIND, Inter. OVERWIND and HF OVERWIND are shown in (b)-(d), respectively. The second row shows the corresponding strains.  The red and blue rectangles in (a) are considered as target and background areas for CNR calculation. }
		\label{phantom_virtual_lateral}
	\end{figure*}  	
\subsection*{Phantom Data}
The phantom  data is acquired from a tissue mimicking breast phantom (059 tissue mimicking breast phantom, CIRS tissue simulation \& phantom technology, Norfolk, VA, USA) using an E-Cube R12 ultrasound machine (Alpinion, Bothell, WA, USA) with a L3-12H probe at the center frequency of $8$ MHz and sampling frequency of $40$ MHz.  The lateral sampling frequency in HF OVERWIND is 8 times higher than LF OVERWIND, therefore, the data is interpolated by a factor of 8 using the cubic spline method for Inter. OVERWIND.   
\section{Results}
\subsection*{Simulation Results}
Fig. \ref{sim_virtual_lateral} shows the lateral displacement and strain for LF OVERWIND, Inter. OVERWIND and HF OVERWIND. It is clear that elastography on data sampled at a higher rate significantly  improves the estimations. As anticipated, the interpolation decreases the variance of estimation significantly at the expense of over smoothing compared to low sampled data.  The reported RMSE, ME and VE and CNR in table \ref{sim_rmse_lateral} also corroborate improvements in lateral estimations. To provide a better comparison, we illustrate the Edge Spread Function (ESF) of the estimated strains across two  vertical and horizontal lines  shown in Fig.~\ref{sim_virtual_lateral} (a). As illustrated in Fig.~\ref{esf_lateral}, the ESF of the HF OVERWIND is substantially closer to the ground truth as compared to those of Inter. OVERWIND and LF OVERWIND.

\begin{table}
	\begin{center}
		\caption{Quantitative comparison of lateral strain estimation on simulated phantom.}
		\label{sim_rmse_lateral}
		\begin{tabular}[c]{cccc}
			\hline		
			&LF&Inter.&	HF\\
				&OVERWIND&OVERWIND&	OVERWIND\\
			\hline		     	     
			ME	&$-1.1\times 10^{-3}$ & $1.6\times 10^{-3}$ &$4.33\times 10^{-4}$ \\
        	VE	& $9.77\times 10^{-6}$&$2.18\times 10^{-6}$  & $6.21\times 10^{-7}$\\
		    RMSE& $85.04\%$& $56.73\%$ & $23.09\%$ \\
			CNR	& 0.37 & -1.49 & 21.70\\				
			\hline
		\end{tabular}\\
	\end{center}
\end{table}

Fig. \ref{sim_virtual_axial} also shows the axial displacement and axial strain for laterally low sampled data, interpolated data and laterally high sampled data. It is inevitable that correct lateral estimation leads to slightly improved axial estimations.  Table \ref{sim_rmse_axial} and Fig. \ref{esf_axial} show marginal improvement of axial strain.   
\begin{table}
	\begin{center}
		\caption{Quantitative comparison of axial strain estimation on simulated phantom.}
		\label{sim_rmse_axial}
		\begin{tabular}[c]{cccc}
			\hline		
			&LF&Inter.&	HF\\
			&OVERWIND&OVERWIND&	OVERWIND\\
			\hline		     	     
			ME	&$-3.13\times 10^{-5}$ & $-3.93\times 10^{-5}$	 & $-4.38\times 10^{-5}$\\
			VE	&$5.21\times 10^{-7}$ & $4.97\times 10^{-7}$ &  $4.96\times 10^{-7}$ \\
			RMSE& $7.99\%$ & $7.82\%$ & $7.81\%$ \\
			CNR	& $50.76$ & $52.56$ & $54.09$\\				
			\hline
		\end{tabular}\\
	\end{center}
\end{table}
\subsection*{Phantom Results} 	
\begin{figure*}
	\centering
	\subfloat[LF OVERWIND]{{\includegraphics[width=4cm]{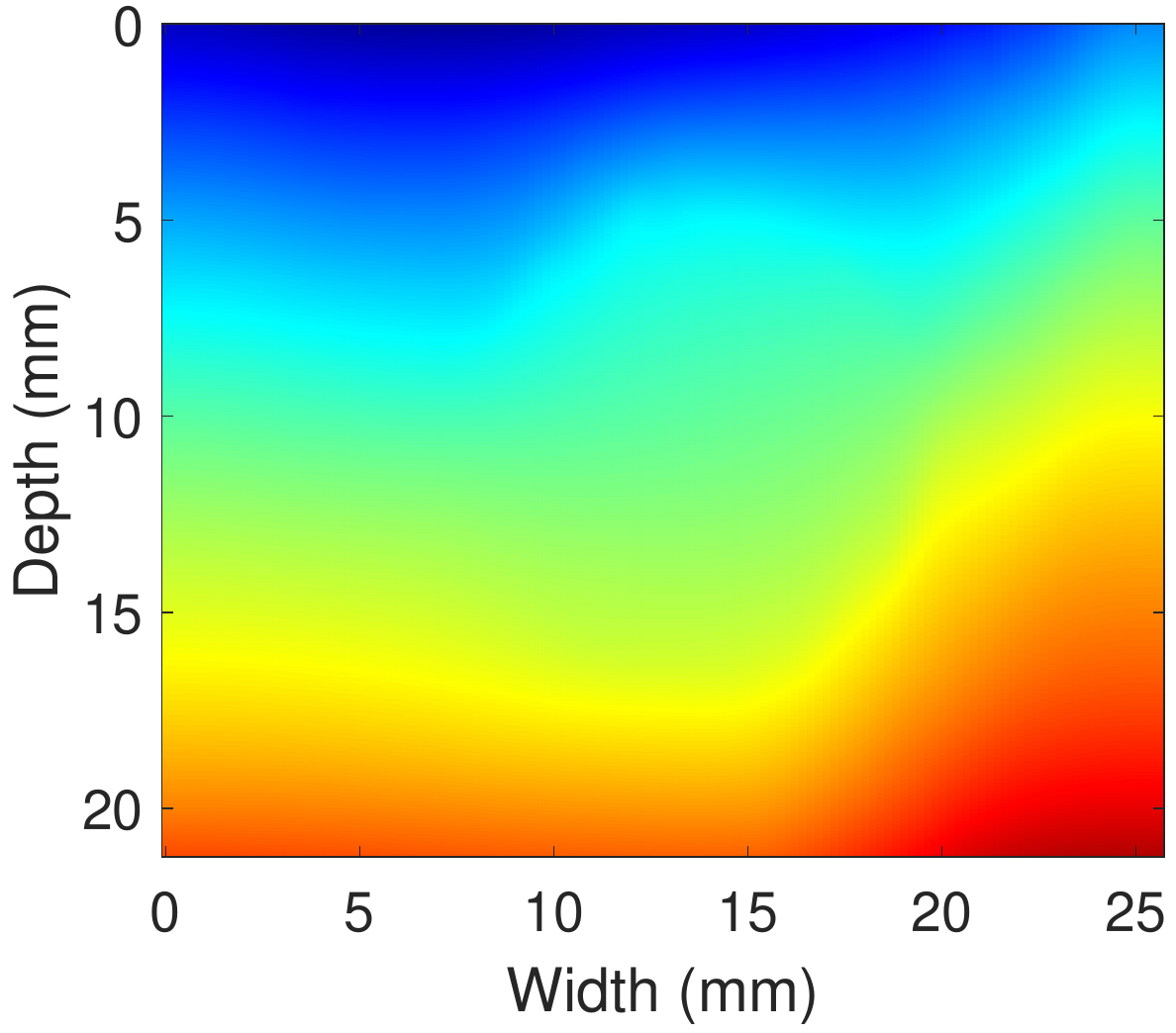}}}
	\subfloat[Inter. OVERWIND]{{\includegraphics[width=4cm]{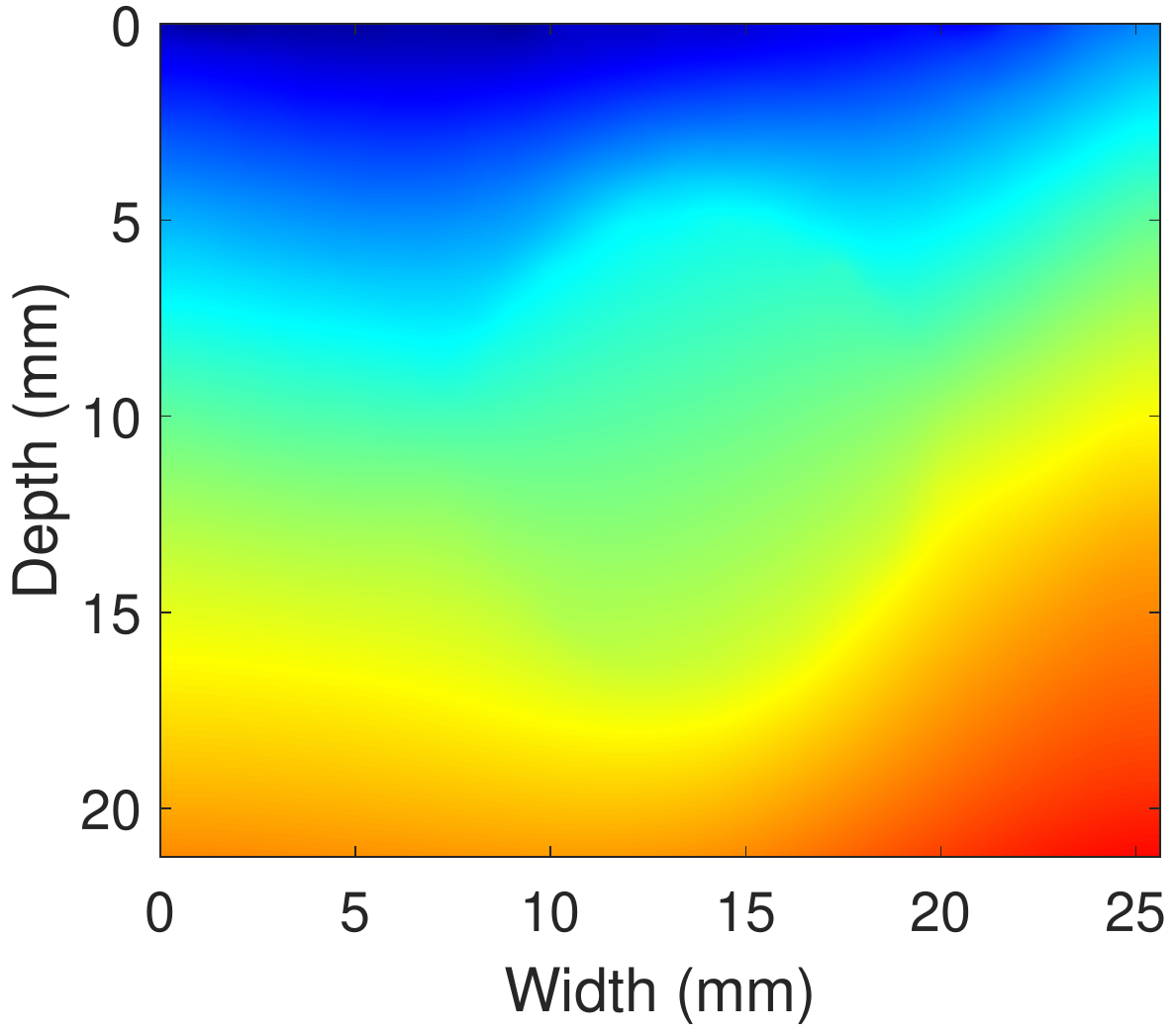}}}
	\subfloat[HF OVERWIND]{{\includegraphics[width=4cm]{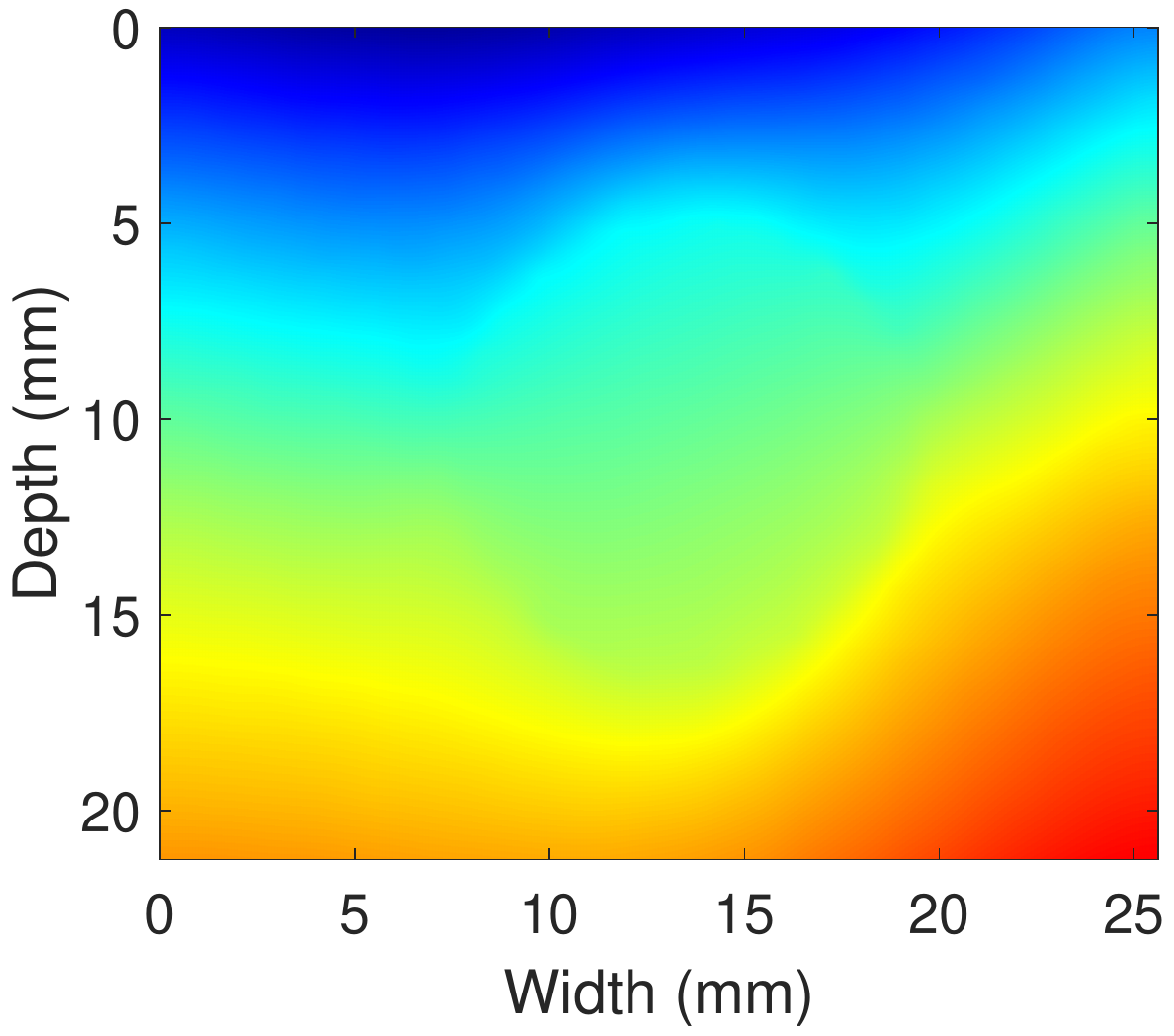}}}
	\subfloat{{\includegraphics[width=0.64cm]{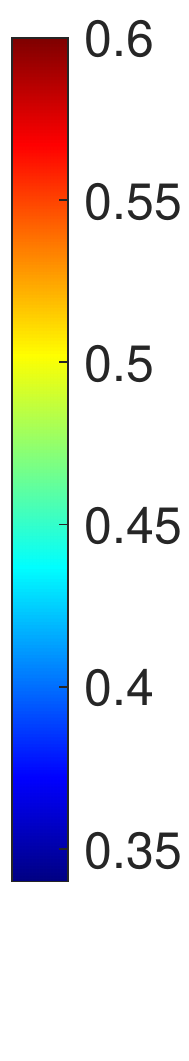}}}
	
	\subfloat[LF OVERWIND]{{\includegraphics[width=4cm]{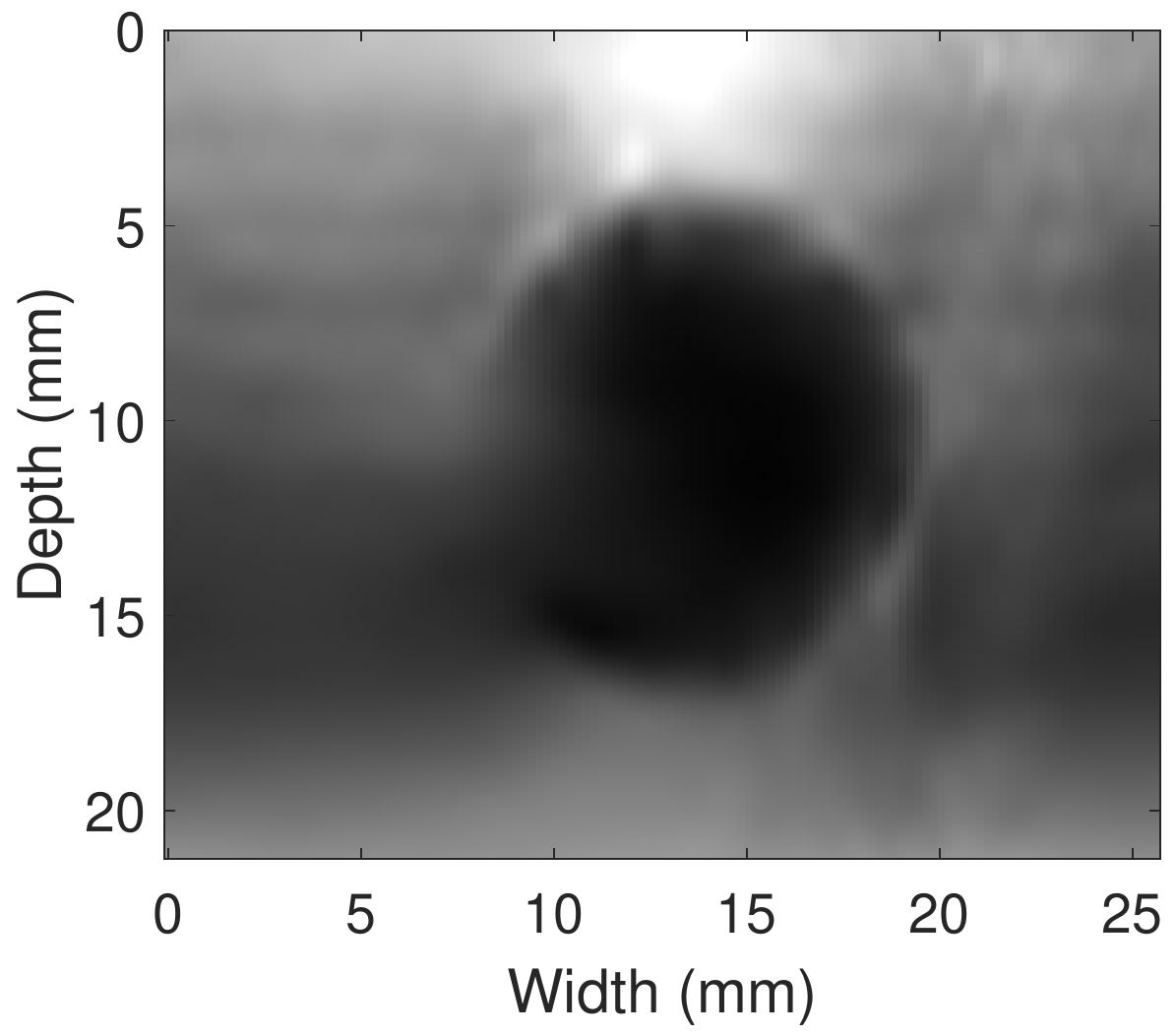}}}
	\subfloat[Inter. OVERWIND]{{\includegraphics[width=4cm]{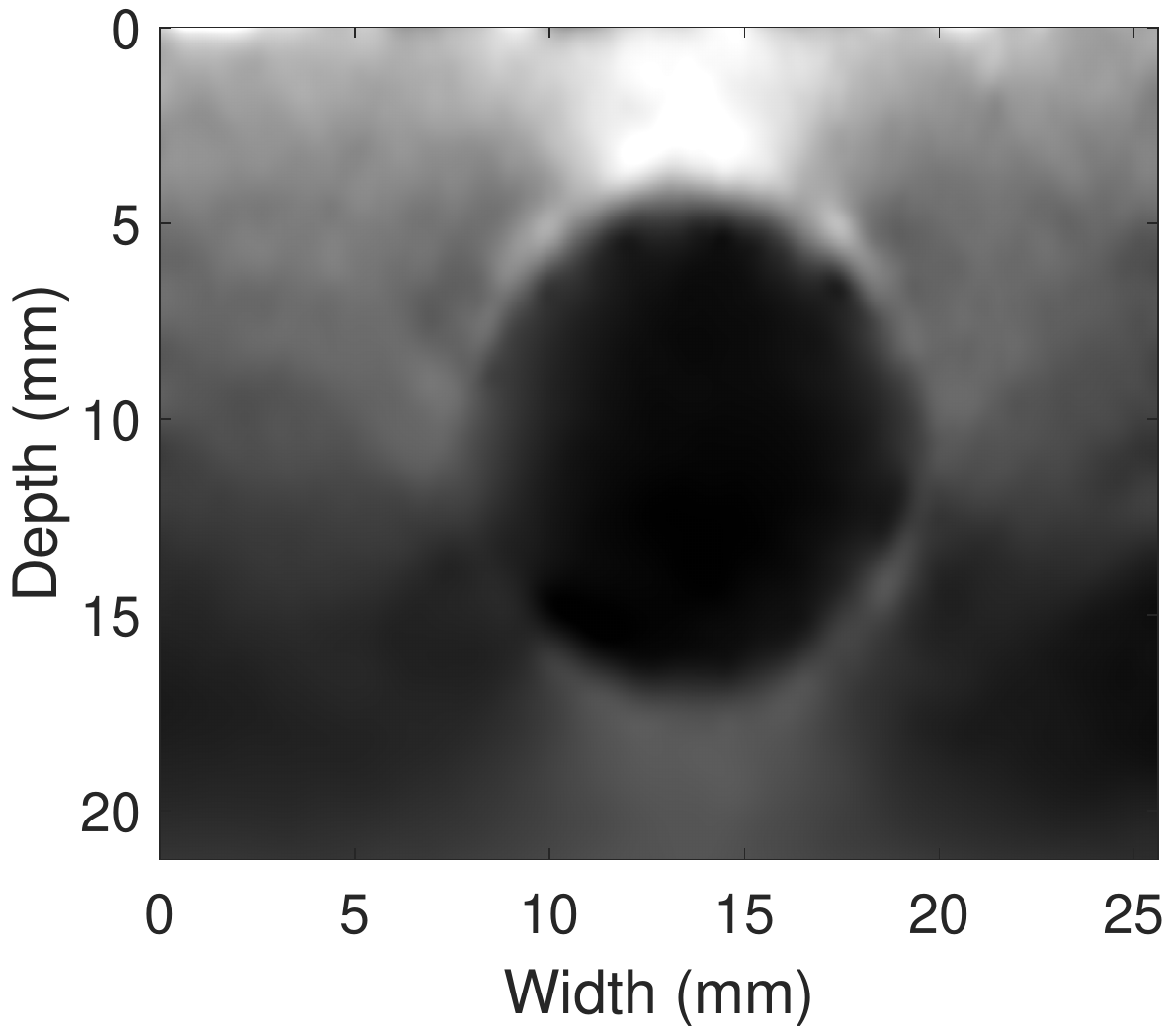}}}
	\subfloat[HF OVERWIND]{{\includegraphics[width=4cm]{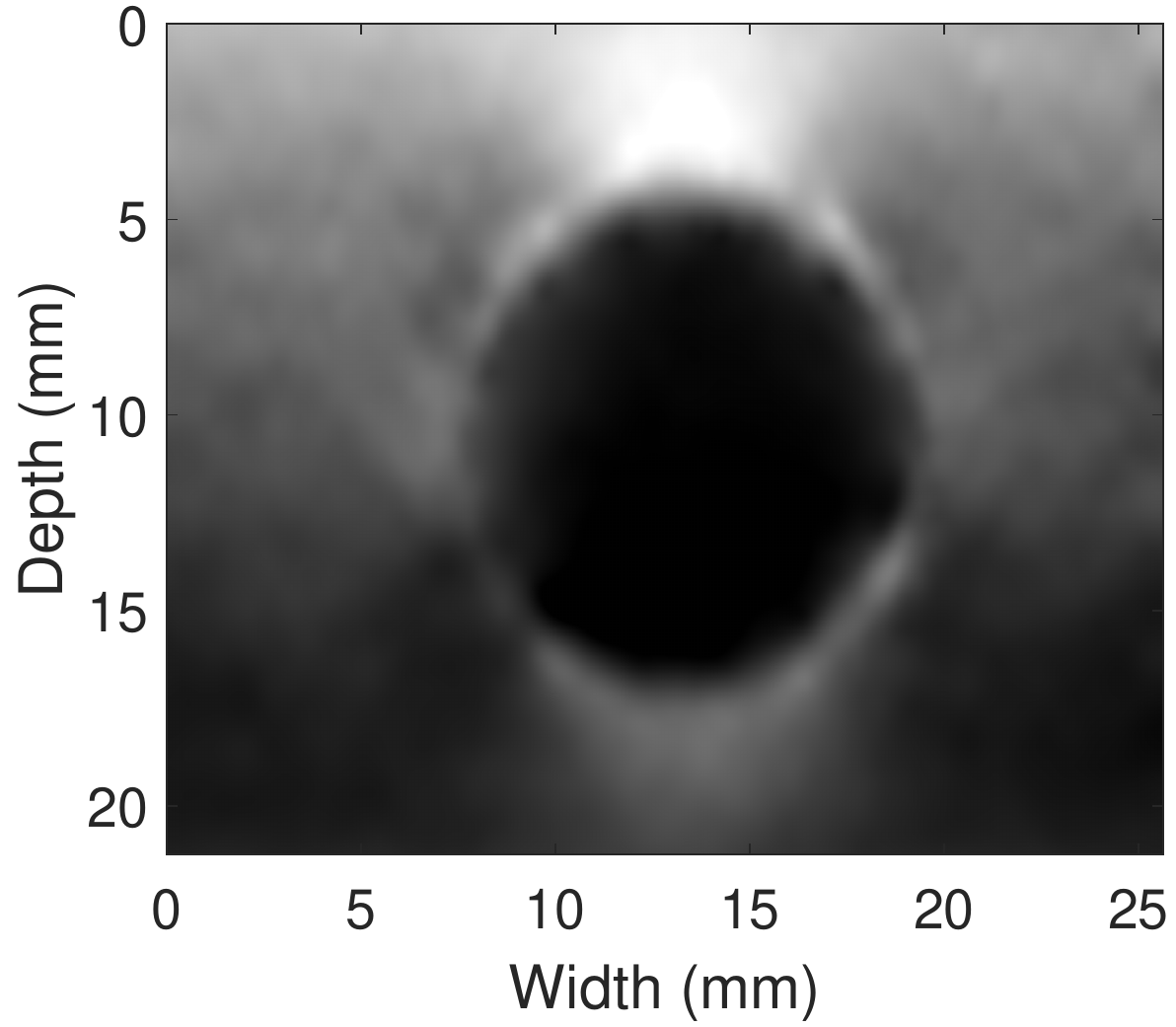}}}
	\subfloat{{\includegraphics[width=0.75cm]{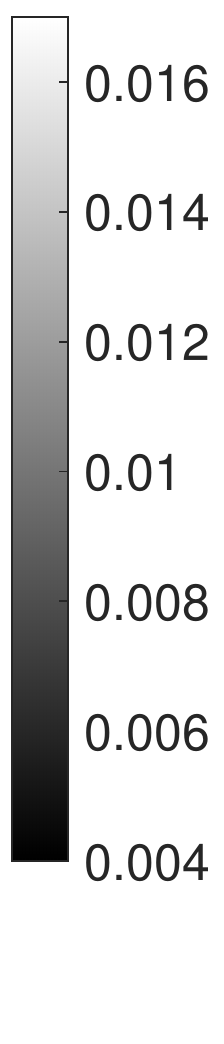}}}
	
	\caption{Results on a tissue mimicking phantom. Estimated axial displacement with LF OVERWIND, Inter. OVERWIND and HF OVERWIND are shown in (a)-(c), respectively. The second row shows the corresponding strains.}
	\label{phantom_virtual_axial}
\end{figure*}

  Estimated lateral displacement and strain for experimental phantom are shown in Fig. \ref{phantom_virtual_lateral}. Similar to the simulation study, the lateral estimation by spline interpolation is over smoothed and HF OVERWIND outperforms the previous methods. Fig. \ref{phantom_virtual_axial} shows the axial displacements and strains and it illustrates better performance of HF OVERWIND over both LF OVERWIND and Inter. OVERWIND. The reported CNR values in Table \ref{CNR_PHANTOM} also show improvement in both lateral and axial estimations by HF OVERWIND.
\begin{table}
	\begin{center}
		\caption{The CNR comparison of different method on the phantom experiment in axial and lateral estimations.}
		\label{CNR_PHANTOM}
		\begin{tabular}[c]{lcc}
			\hline		
			&  \multicolumn{2}{c}{CNR} \\
			\cline{2-3} 
			& Axial & Lateral \\
			\hline
			LF OVERWIND	& 10.80 & -29.89   \\
			inter OVERWIND	&10.28 & -1.55\\
            HF OVERWIND	&11.01 & 6.35\\
			\hline		
		\end{tabular}
	\end{center}
\end{table}

\section{Discussion}
In this paper we proposed to use VSSA as an advanced beamforming technique for time delay estimation in OVERWIND. In this imaging technique multiple piezo-electrics participate in the transmission which improves the beam strength in deep regions. The focused region of the beam can be considered as a virtual element that transmits a  beam in tissue. As such, the imaging procedure becomes closer to the operation of the synthetic aperture and received data can be beamformed similar to synthetic aperture. In this imaging mode, the sampling frequency in the lateral direction can be increased as much as the sampling frequency in the axial direction to increase the resolution and addresses two of the major limitations in estimation lateral displacements. Meanwhile, VSSA has fixed and narrow beam width in all imaging field which results in accurate and high-resolution displacement estimation. 

\begin{figure}[h!]
	\centering
	\subfloat[]{{\includegraphics[width=4cm]{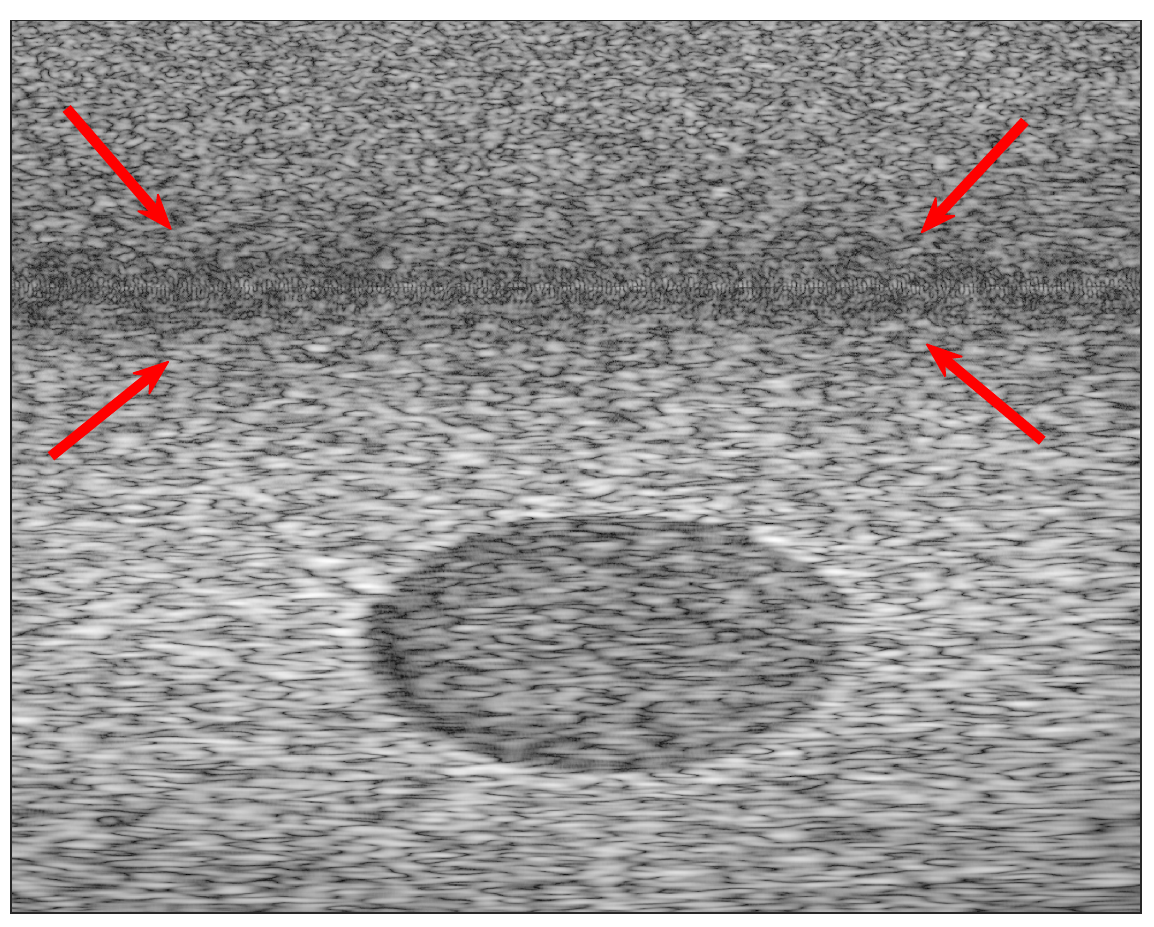}}}	
	\caption{Discontinuity of the VSSA imaging in the focal depth}
	\label{virtual}
\end{figure}

Virtual source synthetic aperture assumes the focal point as a beam source that transmits the signal and conducts beamforming according to Eq. (\ref{virtual_focus}). The $\pm$ term in Eq. (\ref{virtual_focus}) divides the imaging area into top and bottom regions above and below the virtual source resulting in a discontinuity at the focal depth as shown in Fig. \ref{virtual}. This discontinuity is a source of error for USE. Therefore, the focal point should be established in an area outside the region of interest
and corresponding data should be cropped before elastography to avoid this discontinuity.

In addition to the advantages associated with HF OVERWIND with respect to higher accuracy and increased resolution in its estimation, tuning the parameters in HF OVERWIND is much easier than LF OVERWIND. The OVERWIND cost function has four regularization parameters, namely $\alpha_1, \alpha_2, \beta_1$ and $\beta_2$. The parameters $\alpha_1,$ and $\beta_1$ regularize displacements of two neighboring samples in the same A-line in axial and lateral directions, while $\alpha_2, \beta_2$ regularize displacements of two neighbor samples at the same depth and neighbor A-line. As a rule of thumb, by assuming the Poisson's ratio of biological tissues close to $0.5$, lateral displacements of two samples is half of axial displacements, therefore $\beta_1$ and $\beta_2$ can be adjusted as $\beta_1=0.5*\alpha_1$ and $\beta_2=0.5*\alpha_2$, to reduce the number of parameters to two. In HF OVERWIND, the sample size in axial and lateral directions is equal. Therefore  $\alpha_2$ can be set equal to $\alpha_1$ to reduce the number of parameters to one parameter.   
  
 The VSSA yielding high sampling frequency and high resolution in the lateral direction can be utilized with all elastography techniques and it  significantly improves the results. For window-based techniques such as NCC, it is important to note that they are slow and computationally expensive techniques even for data with a low sampling frequency in the lateral direction. In window-based techniques, the running time has a linear relationship with image size which is a disadvantage of these techniques for VSSA with high sampling frequency in the lateral direction. The data with high sampling frequency can also be utilized for other regularized optimization-based elastography techniques. However, over-smoothness is a challenging issue for regularized optimization-based techniques due to a low resolution and also smaller deformation in the lateral direction and it is shown in \cite{overwind} that OVERWIND has a better capability in estimation the sharp transitions.

\section{Conclusion}
Accurate estimation of the tissue mechanical parameters requires accurate strain estimation in all direction. 
Although elastography techniques estimate displacements in both axial and lateral directions, estimation in the axial direction is more accurate than in the lateral direction due to high sampling frequency, improved axial resolution and a carrier signal propagating in the axial direction. 
 In this paper we proposed to use VSSA imaging mode to benefit from advantages of both SA and line by line imaging in a high resolution with high number of A-lines and penetration depth. The beamformed data is fed to our recently developed TDE method, OVERWIND. The results exhibit significant improvement compared to interpolating data in lateral direction as one of the commonly used techniques in estimating lateral strain.
\section*{Acknowledgment}
We acknowledge the support of the Natural Sciences and Engineering
Research Council of Canada (NSERC) RGPIN-2020-04612 and RGPIN-2017-06629.
We thank Alpinion for technical support. 




\FloatBarrier
\bibliographystyle{IEEEtran}
\bibliography{ref}

\end{document}